\documentclass[%
 aip,
 jmp,%
 amsmath,amssymb,
 preprint,%
]{revtex4-2}
\pdfoutput=1
\usepackage{mathrsfs}

\usepackage{graphicx}
\usepackage{dcolumn}
\usepackage{bm}
\usepackage[]{lineno} 

\usepackage{graphicx}
\usepackage{adjustbox}
\usepackage{hvfloat}
\usepackage{ragged2e}           

\DeclareUnicodeCharacter{0301}{*************************************}

\usepackage{dcolumn}
\usepackage{bm}

\usepackage[T1]{fontenc}
\usepackage{mathptmx}
\usepackage[colorlinks = true,
            linkcolor = blue,
            urlcolor  = blue,
            citecolor = blue,
            anchorcolor = blue]{hyperref}
\usepackage{nameref}
\usepackage{flafter}
\usepackage{float}
\usepackage{amsmath, amssymb, latexsym}
\usepackage{mathtools}
\usepackage{color}
\usepackage{tikz}
\usepackage{pifont}
\usepackage{rotating}
\usepackage{enumitem}
\usepackage[T1]{fontenc}
\setlist[enumerate]{itemsep=0mm}
\usepackage{tabularx}
\usepackage{array}
\usepackage{etoolbox}
\usepackage{xcolor,graphicx}
\usepackage{epstopdf, epsfig}
\usepackage{hyperref}
\usepackage{color, soul}   
\usepackage{nomencl}
\usepackage{caption}
\usepackage{natbib}
\usepackage{paralist}
\usepackage{nomencl}
\makenomenclature
\usepackage[left=2.5cm, right=2.5cm, top=3cm, bottom=2.5cm]{geometry}

\usepackage{calrsfs}
\DeclareMathAlphabet{\pazocal}{OMS}{zplm}{m}{n}

\AtBeginDocument{%
  \setlength{\linenumbersep}{\dimexpr\oddsidemargin+1in-\linenumberwidth\relax}%
}

\AtBeginDocument{%
  \setlength{\linenumbersep}{\dimexpr\oddsidemargin+0.1in-\linenumberwidth\relax}%
}
\renewcommand{\makeLineNumber}{\llap{\linenumberfont\rlap{\LineNumber}\hspace{1.3cm}}}

\begin{document}
\mbox{}
 


\title{Transition to turbulence in randomly packed porous media; scale estimation of vortical structures}


\author{Reza M. Ziazi}%
 \email{ziazir@oregonstate.edu}
\altaffiliation[]{}

\author{James A. Liburdy}
 \email{james.liburdy@oregonstate.edu}
\affiliation{%
School of Mechanical, Industrial, and Manufacturing Engineering, Oregon State University, Corvallis, Oregon 97331, USA
}%


\begin{abstract}
\justifying
\linespread{1}\small{
Pore-scale observation of vortical flow structures in porous media is a significant challenge in many natural and industrial systems. Vortical structure dynamics is believed to be the driving mechanism in the transition regime in porous media based on the pore Reynolds number, $Re_p$. To examine this assertion, a refractive-index matched randomly packed porous medium with homogeneously-sized glass spheres ($D_B = 15 \rm mm$) is designed to measure the scale of vortical flow structures in transition from unsteady laminar to turbulent using two-dimensional time-resolved particle image velocimetry. Planar Particle Image Velocimetry (PIV) data is used to quantify the scale of these structures with regard to size, strength, and number density using two different scalings (i) $Re_p$ macroscopic (global), and (ii) $Re_{\langle p\rangle}$ microscopic (local, pore-scale). Data is obtained for $Re_p$ from 100 to 948 in six different locations from the center of the bed towards the wall (one bead diameter away from wall). Direct measurement of vortex scale is quantified by employing swirl strength ($\lambda_{ci}$), vortex core ($\varGamma_{\tiny{2}}$), and enhanced swirl strength ($\lambda_{cr}/\lambda_{ci}$) vortex identification methods. These scales are compared with turbulent integral scales. Due to the confinement of the random medium, the inertia-dependent topology of the flow creates shear-dominant vortical structures in moderate unsteady laminar Reynolds numbers ($Re_p <300$), while the swirl-dominant flow structures appear in weak turbulent Reynolds numbers ($Re_p>500$). From the macroscopic point of view, (i) the size of vortical structures decreases asymptotically to reach 20$\%$ of the global hydraulic diameter, (ii) the strength of vortices increases monotonically by enhancing the inertial effects of $Re_p$, and (iii) the average number density of vortical structures grows from unsteady laminar to fully turbulent. From the pore-scale (local) point of view using $Re_{\langle p\rangle}$, (i) the size of vortical structures decreases monotonically with increasing local Reynolds number, (ii) the strength of vortices rises with $Re_{\langle p\rangle}$, and (iii) the number density of vortical structures for different $Re_{\langle p\rangle}$ is invariant relative to the pore size. These findings suggest pore versus macro-scale coupling exists for the scale of vortical flow structures in the transition regime, albeit the scale variation of pore-scale flow structures with local inertial effects is different from the asymptotic values captured in the macroscopic level with increasing $Re_p$.}
\end{abstract}


\maketitle

\bgroup
\section{\label{sec:Introduction}Introduction\protect}
\vspace{-0.015\textwidth}

\noindent Porous media flows have received increased attention in various natural and industrial applications over the past four decades. Depending upon the conditions, the flow may stay laminar or transition where unsteady chaotic behavior is seen at the pore spaces. Laminar flow in porous media has been extensively investigated in diverse industrial and natural processes. On the other hand, turbulent flows has lately been explored in a wide range of applications that also includes transition regime. This regime is observed in diverse disciplines including (1) near-surface environmental flow such as natural and urban canopies as well as fractured rocks and dams \cite{gayev2007flow,finnigan2000turbulence,dontsov2016tip,emerman1986transport, tomac2018experimental,zolfaghari2017blade,geyer2010measurement, manes2009turbulence,wu2019instability}, (2) catalytic systems \cite{rua2016phenomenological,wehinger2015detailed,lucci2017comparison,sharma1991kinetics,latifi1989use}, (3) nuclear pebble bed reactors \cite{dave2018numerical}, (4) heat transfer enhancement in packed beds \cite{de1972heat,vadasz1999weak,vadasz2000route,vadasz2000subcritical,vadasz1999local,vadasz2015feedback,howell1996combustion,hall1994exit}, (5) dense fluidized beds \cite{lu2018direct, deen2012direct, deen2014review}, (6) flow in biological porous media \cite{khalili2010application, khanafer2008flow}, (7) chromatography columns \cite{hlushkou2006transition}, (8) enhanced oil recovery \cite{hassanipour2012analysis,hassanipour2013simulation}, and (9) thin porous media in paper making industry  \cite{singh2015micro,larsson2018tomographic}. These diverse applications demonstrate the need for further study of the transition whether the flow naturally becomes turbulent or is actively enhanced to increase performance.

The transition to turbulence in porous media may appear in a number of ways; (i) global (macroscopic) versus local (pore-scale) development of flow inertia \cite{suekane2003inertial,johns2000local}, (ii) flow heterogeneity leading to non-uniform distribution of flow in a porous bed \cite{sederman1998structure}, (iii) intrinsic flow instabilities \cite{hill2002moderate,hill2002transition}, and (iv) emergence of coherent topological patterns. These patterns \cite{soulaine2017numerical, hill2001moderate,koch2001inertial,hill2002transition,sederman1998structure,chu2018flow,sahimi2011characterization,stoop2019disorder,anna2013mixing} evolve spatially and temporally with inertial dominance in transition to turbulence and contribute to the change of scales. Furthermore, the complex structural geometry at the pore-level \cite{dullien2012porous,shimizu2017impact,vogel2002topological} results in generation of coherent topological regions such as recirculation zones, jet-like flows, channel-like flows, or impinging flows, as it has been seen by Patil and Liburdy \cite{patil2013turbulent}, and Suekane et al. \cite{suekane2003inertial}, and also regions resembling convergent-divergent nozzles, diffusers, wakes, or stagnant zones inside pores. Swirling structures can develop in these regions, which may evolve into turbulent eddies \cite{wu2007vorticity}. To understand proper scales during transition, it is essential to demarcate the viscous-inertial effects on the spatio-temporal evolution of scales. This process can be seen from a macroscopic (global) point of view (defined in Refs.~\onlinecite{scheidegger1958physics, bear_1972,dullien2012porous}) by neglecting the intrinsic flow characteristics. However, pore-scale (local) investigation of local transition is required to understand the overall transition process in randomly packed beds.

\begin{figure*}
 \centering
 \includegraphics[width=1\textwidth]{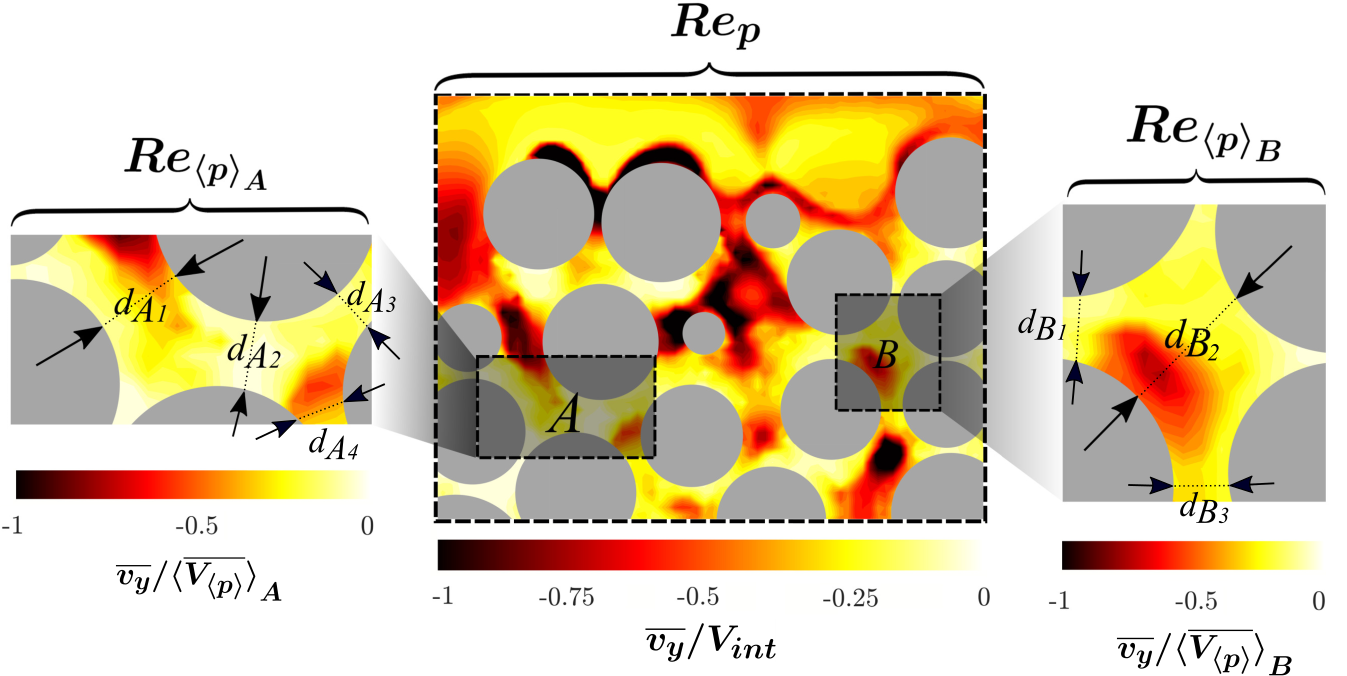}
 \caption{Schematic of the geometrical demonstration of individual pore-scale (local) regions for estimating the local Reynolds number $Re_{\langle p\rangle}$ versus global (macroscopic) for longitudinal (y-direction) velocity intensities normalized by interstitial velocity globally, and by pore-averaged velocity locally. Color bars represent the global and local velocity intensities.}
\label{fig:1}
 \end{figure*}

Reynolds numbers have had different definitions among porous media community (Wood et al. 2020 \cite{wood2020modeling}). These are listed as (i) particle diameter and superficial velocity \cite{jolls1966transition,ward1964turbulent,wright1968nonlinear,wegner1971visual, karabelas1971use,latifi1989use,dybbs1984new, seguin1998experimentaltransition,seguin1998experimentallaminar,bu2014experimental,wehinger2015detailed,thaker2019piv}, (ii) pore hydraulic diameter and interstitial velocity \cite{patil2012optical,patil2013turbulent,patil2015scale,hellstrom2010laminar,finn2013relative,wood2015comparison,khayamyan2017measurements,nguyen2018time,he2018angular,larsson2018tomographic,he2019characteristics,nguyen2019experimental}, (iii) approximate permeability and average intrinsic velocity \cite{kececioglu1994flow,kaviany2012principles}, (iv) pore-averaged velocity and particle size \cite{koch2001inertial,hill2001moderate,hill2002transition,hill2002moderate,reynolds2000flow,horton2009onset,agnaou2016steady,jin2015numerical,chu2018flow}, and (v)  pore-averaged length scale, and pore-averaged velocity. For the latter, the length scale has been based on the square root of pore area (Sederman et al. 1998 \cite{sederman1998structure}), cross-stream length of velocity profile (Johns et al. 2000 \cite{johns2000local}), or along a line perpendicular to the flow direction (Suekane et al. 2003 \cite{suekane2003inertial}). Use of a pore-averaged velocity with some measure of particle size has been applied extensively in regular packed beds \cite{jolls1966transition,reynolds2000flow,hill2002transition,hill2001moderate,hill2002moderate} to perform comparisons with flow over sphere as a canonical case \cite{nakamura1976steady,natarajan1993instability,taneda1956experimental,goldburg1966transition,kim1990stability,johnson1999flow,gumowski2008transition,szaltys2012nonlinear,skarysz2018experimental}. However, this method overestimates actual pore sizes in random packings. 

In this study, two Reynolds numbers are used to investigate the scale characteristics for pore-scale versus macroscopic perspectives. The global Reynolds number ($Re_{p}$) uses method (ii) above and is defined as:
\begin{equation}
Re_{p} = \frac{V_{int}D_{H}}{\nu}
\label{eq:global} 
\end{equation}

\noindent where, $V_{int}$ is the \textit{interstitial} velocity defined as a function of \textit{superficial} or \textit{seepage} velocity (also known as Darcy velocity) as $V_{int} = V_{Darcy}/\Phi$, $D_{H}$ is the hydraulic diameter defined based on the bead diameter and porosity as $D_{H} = \Phi D_{B}/(1-\Phi)$, and $\nu$ is the kinematic viscosity of the fluid. The local pore-scale Reynolds number $Re_{\langle p\rangle}$ is defined here to compensate for the deficiencies of applying hydraulic diameter as the length scale of each arbitrary pore in estimating the actual pore-scale flow inertia and to better define local pore-scale velocities. Hence,

\begin{equation}
Re_{\langle p\rangle} = \frac{ \langle \overline{ V_{\langle p\rangle}}\rangle  \langle d_{\langle p\rangle} \rangle}{\nu}
\label{eq:local} 
\end{equation}

\noindent where, (\textbf{---}) indicates temporal averaging, $\langle \rangle$ represents spatial averaged over a single pore, such that $\langle \overline{ V_{\langle p\rangle}}\rangle$ is the temporally and spatially averaged velocity for a single pore the entire fluid phase as shown in \hyperref[fig:1]{Figure~\ref*{fig:1}}, $\langle d_{\langle p\rangle} \rangle$ is the local pore length scale spatially averaged in the fluid phase in ten locations of pores perpendicular to the local velocity profile (as shown in \hyperref[fig:1]{Figure~\ref*{fig:1}}), and $\nu$ is kinematic viscosity of fluid.  
 
 Investigating the macroscopic scaled flow physics in porous media during transition regimes has led to transition predictions in the literature \cite{ergun1952fluid,ward1964turbulent,dudgeon1966experimental,kyle1971experimental,karabelas1971use,fand1987resistance,kececioglu1994flow,seguin1998experimentaltransition,hlushkou2006transition, hellstrom2010laminar}. These studies have investigated the transition to turbulence with considerable differences in their finding for the range of transition. These estimates of the onset of turbulence are based on the premise of interpreting a micro-scale phenomena with macroscopic parameters (pressure, hydraulic conductivity, macroscopic dispersion, permeability, and mass transfer) or models (Darcy-Forchheimer) \cite{horton2009onset}. 

 Many studies have considered microscopic flow investigation using highly resolved experimental observations, or numerical simulations to identify flow structures for detecting the pore-scale flow physics during the transition, as is shown in \hyperref[fig:3]{Figure~\ref*{fig:3}}. These studies predict the following regimes: (i) steady laminar ($29<Re_p<193$), (ii) unsteady laminar ($46<Re_p<156$), (iii) steady inertial ($1<Re_p<148$), (iv) unsteady inertial ($148<Re_p<300$), (v) viscous-inertial transition ($78<Re_p<480$), and (vi) turbulent ($159<Re_p<727$). The deviations between these predictions are very wide \cite{mickley1965fluid,jolls1966transition,dudgeon1966experimental,wright1968nonlinear,kyle1971experimental,latifi1989use, rode1994hydrodynamics, seguin1998experimentallaminar,seguin1998experimentaltransition, bu2014experimental}. The disparity of these predictions provides motivation to study the entire regime ($100<Re_p<948$) to identify the scale of vortical structures in this range for a different perspective of scale evolution in transition.  Non-invasive imaging techniques to measure local velocity in porous media have been applied for several decades, with high resolution capabilities recently developed.\\
 \hvFloat[
 floatPos=!htbp!,
 capWidth=h,
 capPos=after,
 capAngle=90,
 objectAngle=90,
 capVPos=c,
 objectPos=c,
 nonFloat=true]{figure}{\includegraphics[width=1.35\textwidth]{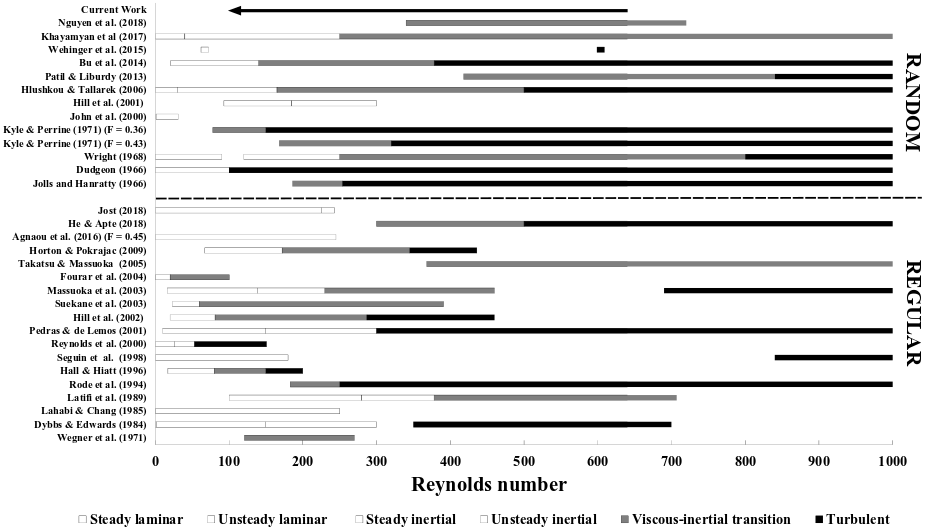}}
{Regime map for pore-scale flow through random and regular porous media during transition for variations of global pore Reynolds numbers (see Refs.~\onlinecite{jolls1966transition,dudgeon1966experimental,wright1968nonlinear, suekane2003inertial, wegner1971visual,kyle1971experimental,dybbs1984new,lahbabi1985high, johns2000local, latifi1989use,rode1994hydrodynamics,seguin1998experimentallaminar,seguin1998experimentaltransition, reynolds2000flow,hill2001moderate,hill2002transition,masuoka2003chaotic,fourar2004non,takatsu2005transition,hlushkou2006transition,horton2009onset,patil2013turbulent,bu2014experimental,agnaou2016steady,khayamyan2017measurements,he2018angular, nguyen2018time,jost2018direct, wehinger2015detailed, hall1994exit}).}{fig:3}\\
%

Methods that have been applied include particle image velocimetry (PIV) \cite{saleh1992measurement,northrup1993direct,masuoka2003chaotic,patil2013turbulent,patil2014experimental, wood2015comparison,kazemifar2016quantifying, khayamyan2017measurements,nguyen2018time, ziazi2019vortical, larsson2018tomographic}, particle tracking velocimetry  (PTV) \cite{peurrung1995measurement,moroni2001statistical,lachhab2008particle}, Laser Doppler Anemometry (LDA)\cite{dybbs1984new,yarlagadda1989experimental}, Magnetic Resonance Imaging (MRI) \cite{johns2000local,elkins2007magnetic,suekane2003inertial,gladden2003magnetic}, Positron Emission Tomography (PET) \cite{khalili1998flow}, and X-ray imaging \cite{soulaine2016impact}.
 
 The studies mentioned in \hyperref[fig:3]{Figure~\ref*{fig:3}} were mostly focused on predicting the onset of turbulence to help reveal the flow physics during transition regime. The current study is aimed to investigate the scale development of flow structures during transition using Eulerian PIV methods without necessarily delineating the onset of turbulence. The vortical structures are one of the major mechanisms during the transition regime that are influenced by mean flow characteristics in development, stretching, and breakup processes all which can alter the scales of motion \cite{wu2007vorticity, tennekes1972first}.

\begin{figure*}
 \centering
 \includegraphics[width=0.6\textwidth]{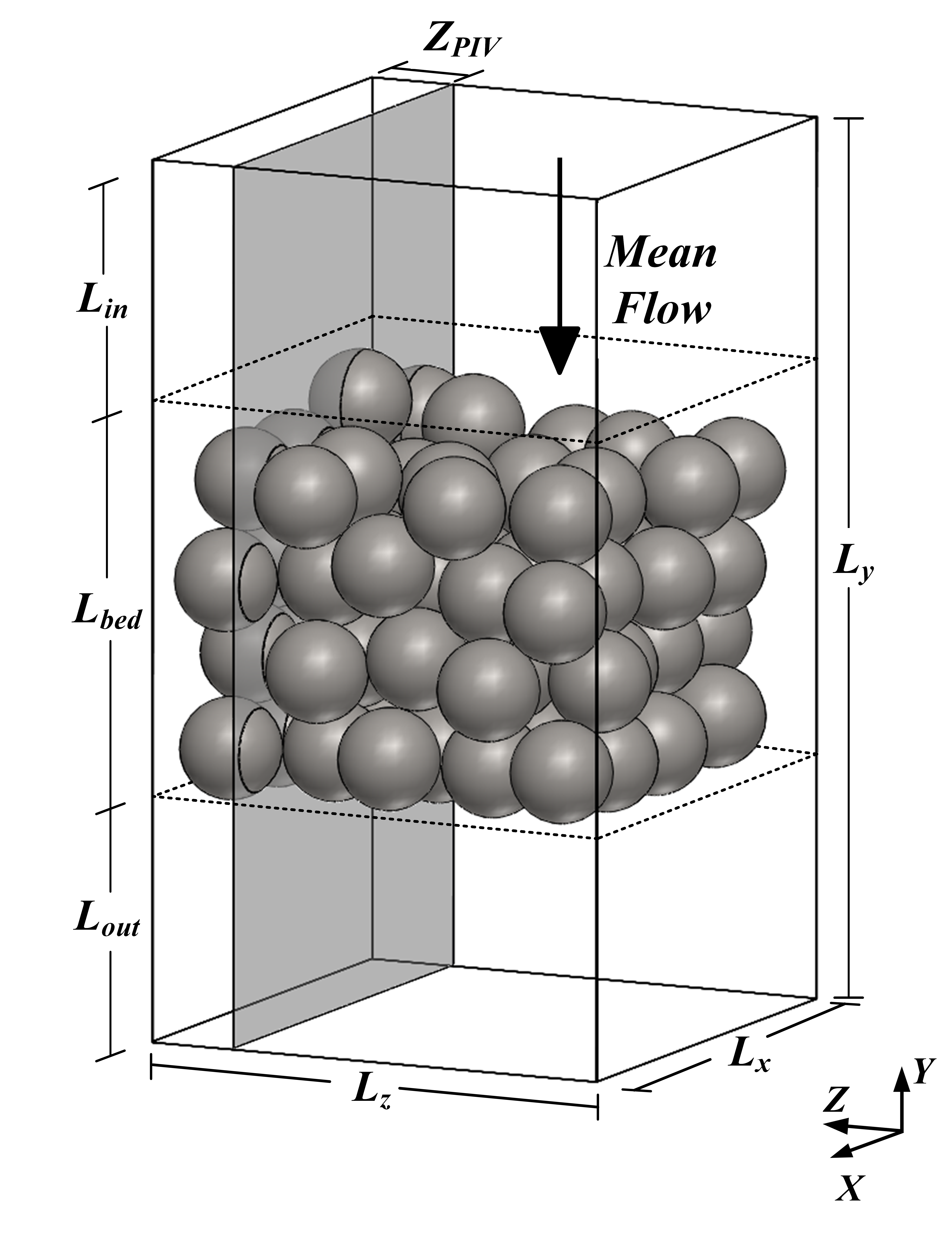}
 \caption{The three-dimensional geometry of the randomly packed porous bed. The location of one of the image planes for PIV data collection is shown for reference, image axis is along coordinate Z.}
\label{fig:bed}
\vspace{-1.5em}
 \end{figure*}

 Refractive Index Matching (RIM) is required for optical imaging has been successful in a few studies \cite{saleh1992measurement,northrup1993direct,moroni2001statistical,arthur2009piv, patil2012optical,nguyen2018time}. In the present work, high fidelity time-resolved PIV technique is used to measure the velocity field based on RIM. The results are used to address the following:

\begin{compactenum}[(i)]
    \item determination of the scaling of size, strength, and number density of the local vortical structures based on global and local pore scaling.
    \item evaluation the dominant mechanisms, such as rotation or shear characteristics of vortices that influence the scale of the vortical structures in transition.
    \item identification asymptotic limits for the size, strength, and number density of vortical structures in transition using local and global scaling.
\end{compactenum}

This paper is organized as follows. In \hyperref[sec:methodology]{Section~\ref*{sec:methodology}}, the details of geometry of the porous medium, flow-loop setup, measurement parameters, experimental method, and uncertainties are provided. Experimental results are presented in \hyperref[sec:results]{Section~\ref*{sec:results}} with emphasis on the evolution of vortical structure size, scale, and number density during transition using vortex identification methods. In \hyperref[sec:conclusion]{Section~\ref*{sec:conclusion}} a summary discussion of the results together with conclusions are given. 

\section{Methodology}\label{sec:methodology}
The porous medium was designed for the collection of high-resolution velocity data with high quality images with low velocity uncertainty. In doing this the following elements were considered: (i) the geometry of porous system, (ii) refractive-index matching of porous medium, (iii) experimental setup detailing the method for collecting PIV velocity using RIM, and (iv) the uncertainty quantification for the PIV system.

\begin{figure}
 \centering
 \includegraphics[width=0.9\textwidth,decodearray=1 0 1 0 1 0]{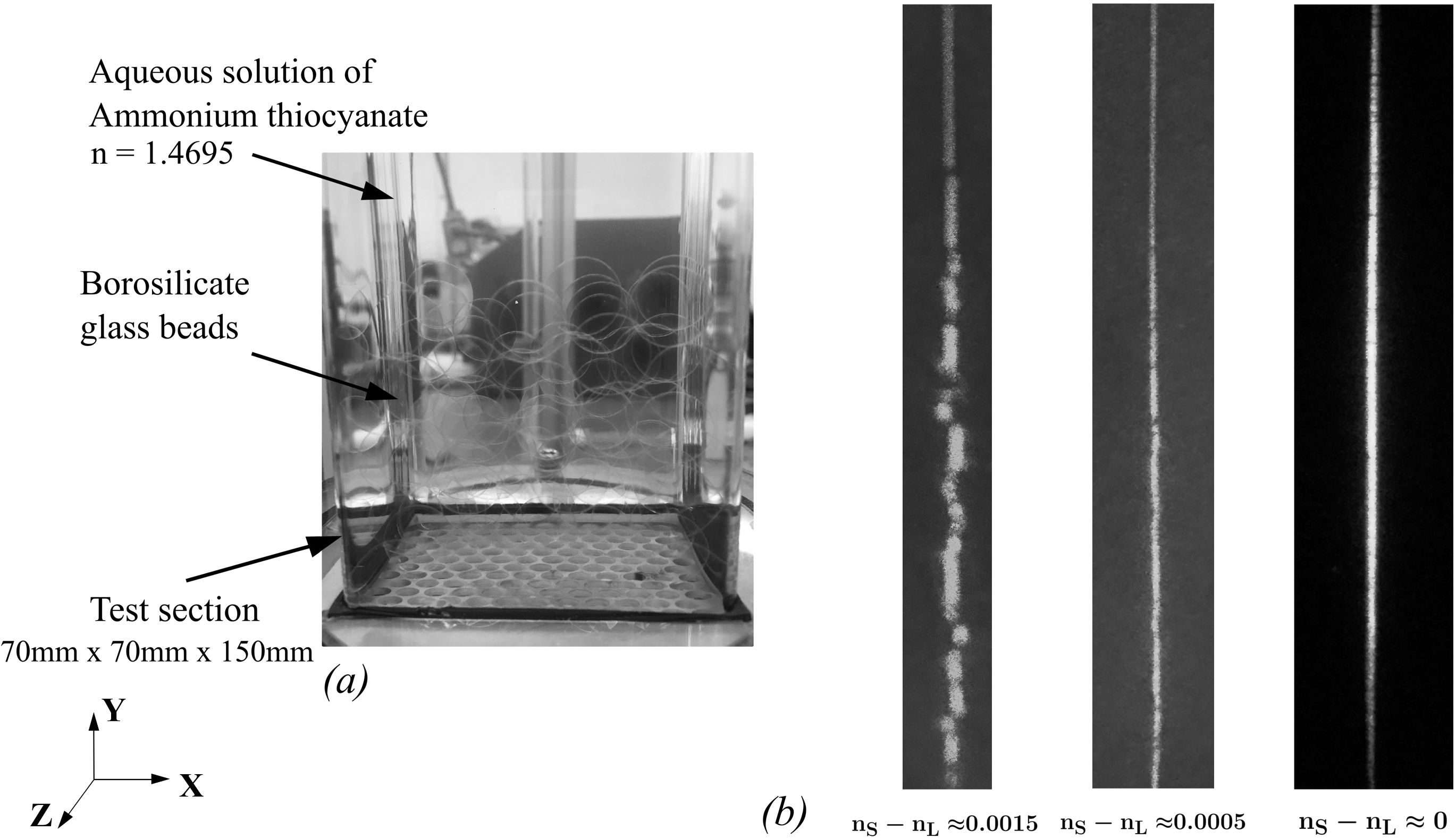}
 \caption{(a) Porous medium with solid and fluid phase refractive indices matched, (b) sequence of images of laser light sheet distortion when traveling through an index mismatched porous bed for light-sheet wavelength of $\lambda$ = 527 \textit{nm} by applying these index mismatch values. }
\label{fig:RIM}
 \end{figure}

\subsection{Geometry of porous medium}\label{subsec:geom}
The random packed geometric arrangement is illustrated schematically in \hyperref[fig:bed]{Figure~\ref*{fig:bed}}. The plane of interest for PIV measurements is capable of being moved to any position along the optical axis (z-direction). The spherical beads, 15 $\rm mm$ diameter made from borosilicate glass, are optically homogeneous and transparent, with less than 0.01$\%$ uncertainty in their diameter. The mean flow convective velocity direction is from top to bottom in the negative Y-direction. Flow enters upstream of the porous bed through a channel with the same cross-section as the porous bed. Details of the test section, beads, and fluid are given in \hyperref[tab:Flowspec]{Table~\ref*{tab:Flowspec}}.

\begin{figure*}
 \centering
\includegraphics[width=1\textwidth]{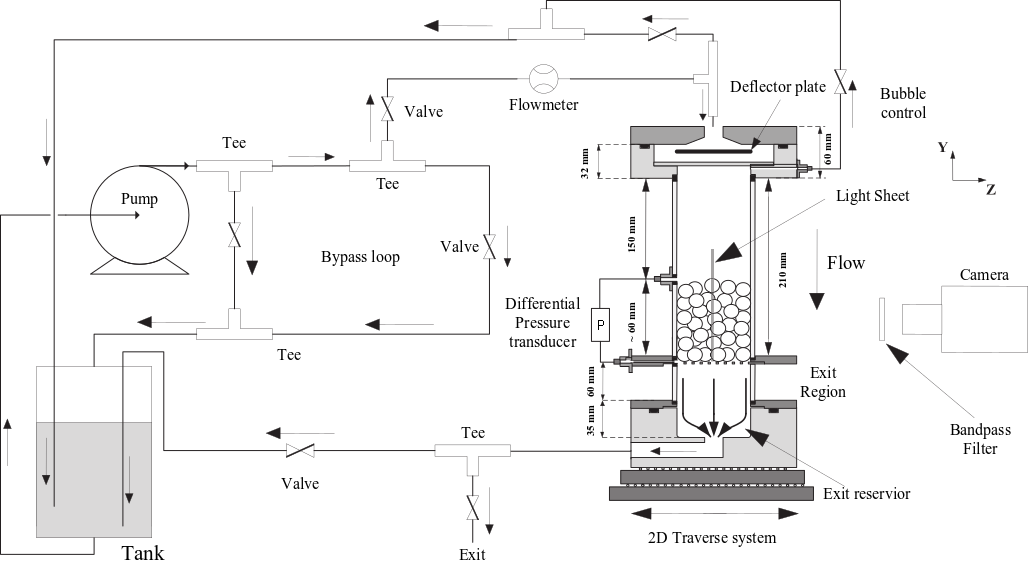}
 \caption{Schematic of the experimental flow loop and test section for PIV imaging together with the bubble control, and the bypass loop}
\label{fig:loop}
 \end{figure*}

\subsection{Refractive-index matching}\label{subsec:RIM}
It is necessary to have an optically uniform porous bed for accurate RIM-PIV results \cite{patil2012optical,northrup1993direct,budwig1994refractive,wiederseiner2011refractive,huang2008optical,borrero2016aqueous,hafeli2014piv,wright2017review,hassan2008flow,goharzadeh2005transition,agelinchaab2006velocity}. The effect of even slight mismatches of refractive index has been evaluated using similar beads and fluid \cite{patil2012optical}, with uncertainties associated with possible distortion presented in \hyperref[subsec:uncertainty]{Section~\ref*{subsec:uncertainty}}. As a qualitative form of RIM distortion, \hyperref[fig:RIM]{Figure~\ref*{fig:RIM}\ (a)} shows the porous bed with the random pack of glass spheres with refractive index of $\rm n_S = 1.47$ that is matched with that of the fluid (aqueous solution of ammonium thiocyanate $\rm NH_{4}SCN$) with a refractive index of $\rm n_L = 1.4695$. This corresponds to the middle laser sheet shown in \hyperref[fig:RIM]{Figure~\ref*{fig:RIM}\ (b)}, where there is a 0.03$\%$ mismatch with glass spheres. The laser sheet on the right shows no detectable mismatch between the solid and liquid, while the one on the left has 0.1 $\%$ of mismatch. In this study, the PIV imaging was obtained with the liquid refractive index within the range of $\rm 1.4695<n_L<1.47$ \cite{patil2012optical,patil2013flow,borrero2016aqueous}. The liquid refractive index was measured using a referactometer (Atago co., Model: R5000). 


\begin{table}[htb]
\begin{center}
\caption{Experimental parameters used for the present study in the random porous bed. }
\label{tab:Flowspec}
\begin{tabular}[c]{@{\centering\arraybackslash}p{0.64\linewidth}@{\centering\arraybackslash}p{0.18\linewidth}@{\centering\arraybackslash}p{0.18\linewidth}}
\hline\hline
\textbf{Parameter}&\textbf{Symbol}&\textbf{Value}\\
\hline
\text{Porosity of porous media}& $\Phi$&  0.49\\
\text{Bead Diameter }~$\rm [mm]$& $\rm D_B$ &$15$ \\
\text{Porous bed aspect ratio}& $L_{x}/\rm  D_B$ &$4.67$ \\
\text{Inlet section length, ~{$\rm [mm]$}}& $L_{in}$& $155$\\
\text{Outlet section length, ~{$\rm [mm]$}}& $L_{out}$& $60$\\
\text{Porous medium length, ~{$\rm [mm]$}}& $L_{bed}$& $60$\\
\text{Test section width, $x-$direction~{$\rm [mm]$}}& $L_x$& $70$\\
\text{Test section width, $z-$direction~{$\rm [mm]$}}& $L_z$& $70$\\
\text{Test section height, $y-$direction~{$\rm [mm]$}}& $L_y$& $270$\\
\text{Apparatus glass wall refractive index (at $\lambda = 527~ \rm{nm}$)}& $\rm n_S$&$1.47$\\
\text{Glass Bead Refractive Index (at $\lambda = 527~ \rm{nm}$)}& $\rm n_S$&$1.47$\\
\text{Liquid Refractive Index (at $\lambda = 527~ \rm{nm}$)}&$\rm n_L$ &$1.4695\pm0.0001$ \\
\text{Liquid Kinematic Viscosity}~$\rm [m^2/s]$ & $\nu$ &1.288 $\times$ $10^{-6}$\\
\text{Liquid Density}~[$\rm Kg/m^3$]& $\rho$&1118\\
\text{Tracer Diameter}~[$\rm \mu m$]& $D_{tracer}$&$10 \pm 2$\\
\text{Tracer Density}~[$\rm g/cm^3$]& $\rho_{tracer}$&$2.5 \pm 0.1$\\
\text{PIV conversion ratio}& pixel/mm&18.207\\
\text{PIV resolution}~[$vector/\rm D_{B}$]& $\rm \Delta_{PIV}$&$17.1$\\
\hline\hline
\end{tabular}
\end{center}
\end{table}

\subsection{Experimental setup}\label{subsec:setup}

Two-dimensional PIV images in different x-y planes were taken in discrete locations along the z-direction, the optical axis, shown in \hyperref[fig:bed]{Figure~\ref*{fig:bed}} using a traverse system of the test-section demonstrated schematically in \hyperref[fig:loop]{Figure~\ref*{fig:loop}}. The experimental flow loop (\hyperref[fig:loop]{Figure~\ref*{fig:loop}}) allows for a range of flow rates, and therefore Reynolds numbers. A total of 95 beads of 15 mm diameter, $D_{B}$, were used to create the porous bed. The packing was randomly set by the spheres into the square test section. The bed porosity, $\Phi$ was directly measured volumetrically to be 0.49, and the bed aspect ratio (width to bead diameter ratio) is 4.67. 


The fluid density was 1118 $\rm kg/m^3$ and the tracer particles were manufactured by Dantec Dynamics, with a size of $10 \pm 2 \mu m$. The tracer particle response time was calculated to be much smaller than the Kolmogrov time scale \cite{elghobashi1991particle}, with a Stokes number $St_{tracer} = \rho_{tracer}D_{tracer}^2V_{int}/18 \mu D_{B} \ll 1$, where $D_{tracer}$ is the tracer particle diameter and $\rho_{tracer}$ is the density of tracer particles.
\begin{table}
\begin{center}
\caption{PIV configuration of different cases in the current study}
\label{tab:cases}
\centering

\begin{tabular}[c]{@{\centering\arraybackslash}p{0.05\linewidth}@{\centering\arraybackslash}p{0.25\linewidth}@{\centering\arraybackslash}p{0.35\linewidth}@{\centering\arraybackslash}p{0.25\linewidth}@{\centering\arraybackslash}p{0.1\linewidth}}
\hline\hline
\centering Cases&\centering $Z_{PL}/L_{z}$ &\centering $Re_p$&$\langle \overline{V} \rangle/ V_{int}$ & $\Phi_{PL}$\\
\hline
\centering\text{PL-1}&  \centering 0.5 & \centering 100& 0.82 & 0.45\\
\centering\text{PL-2}& \centering 0.41& \centering 267& 0.86 & 0.46 \\
\centering\text{PL-3}& \centering 0.4& \centering 272&  0.88 & 0.49\\
\centering\text{PL-4}& \centering 0.39& \centering 270&   0.87& 0.46\\
\centering\text{PL-5}&\centering  0.3& \centering 253 &  0.81 & 0.41\\
\centering\text{PL-6}&\centering  0.21& \centering 328-948&  0.82-0.92 & 0.45\\
\hline\hline
\end{tabular}

\end{center}
\end{table}

The imaging was based on high speed laser-pulsed lighting from a Nd-YLF laser (10000 Hz) at nominal wavelength of 527 nm (New Wave Research, Pegasus PIV). The camera was a CMOS (Integrated Design Tools Inc., Model: $\rm MotionPro^{TM}$X-3) with an adjustable focusing lens  (Nikon AF Micro-NIKKOR 60mm f/2.8D). All data were gathered using an optimal f-number setting of 8 to compensate for both resolution and low uncertainty of in-plane motion. The imaging plane did not move relative to camera, hence, the magnification associated for each measuring plane was approximately constant, varying about 8$\%$ along the optical axis for the different planes; from 0.209 to 0.218. The interrogation window was $32 \times 32$ pixels with 50$\%$ overlap to decrease noise. Thereby, the results are in a velocity vector spacing of 0.87 mm, equivalent to 17.1 vectors per bead diameter. Depending on the value of $Re_p$, the sampling frequency for each vector field ranged from 62 to 185 Hz corresponding to the lowest and highest $Re_p$, respectively. A total of 3200 velocity samples for each Reynolds number were captured. A high accuracy iterative adaptive PIV scheme specific to reducing out-of-plane loss of image pairs using $\rm Dantec \ Dynamics^\circledR$ was employed to perform image correlation, adaptive subpixel peak fitting, moving averaging, and subpixel deformation of the interrogation area. The validation process rejected less than $6.4\%$ of a total 4697 vectors as outliers. 

\begin{figure*}
 \centering
 \includegraphics[width=0.9\textwidth]{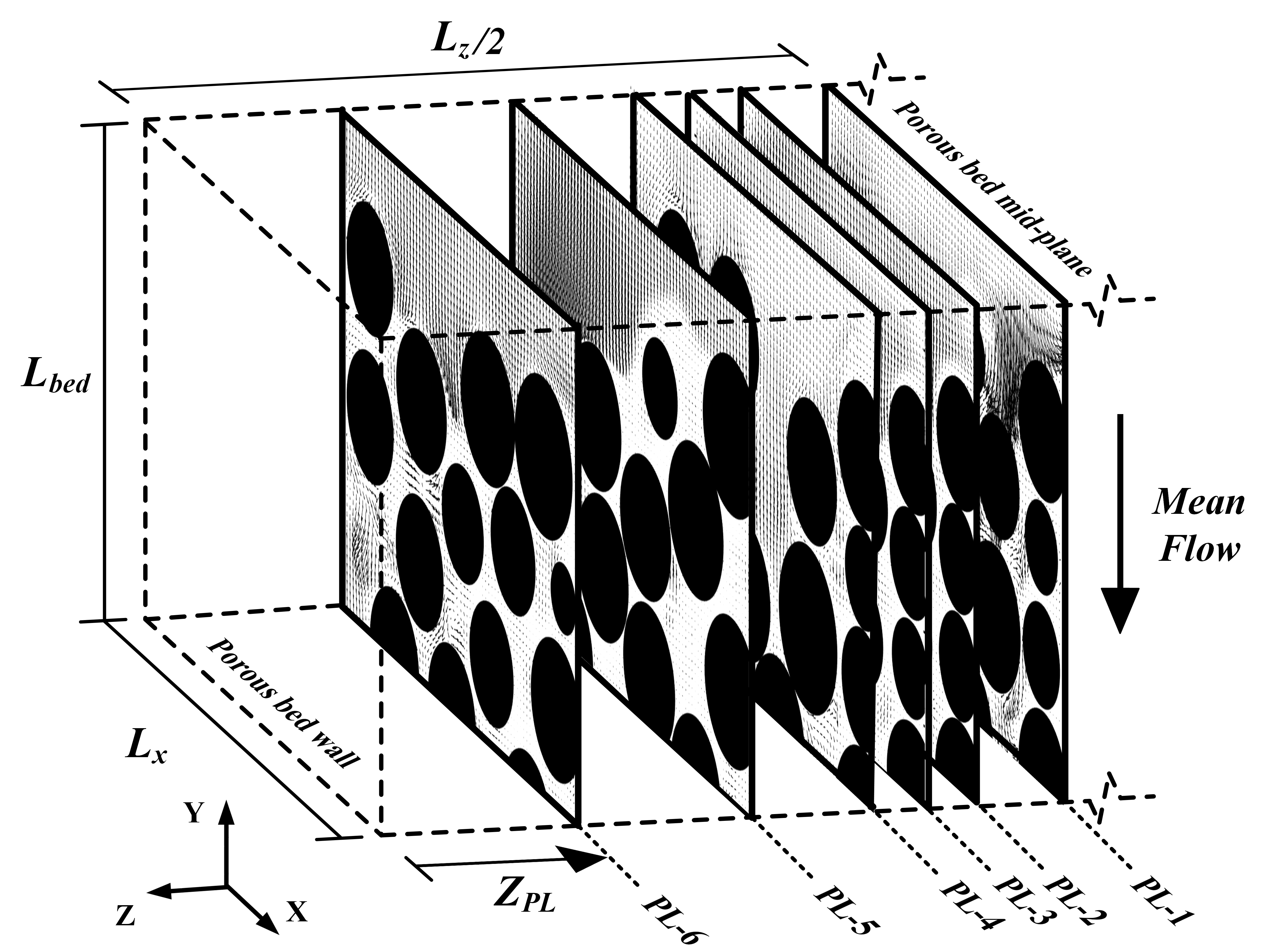}
 \caption{ The vector-map configuration of instantaneous velocity at different planes in the experimental apparatus for Reynolds numbers ranging from 100 to 948. Six different test cases were obtained. }
  \vspace{-1.5em}
\label{fig:8}
 \end{figure*}

 The data were obtained in six different planes along the optical axis (Z-direction) for the pore Reynolds numbers ranging from $Re_{p}=100$ to $Re_p = 948$. \hyperref[fig:8]{Figure~\ref*{fig:8}} is a schematic representation of six planes starting from the middle of the test-section (PL-1) towards the wall (PL-6). The plane closest to the wall was one bead length from the test-section wall. The details of the global conditions in each plane are summarized in \hyperref[tab:cases]{Table~\ref*{tab:cases}}. The size and shape of the pores in PL-2 to PL-4 are very similar with very nearly the same values of $Re_p$ of approximately 270. PL-6 is located within one bead diameter away from the wall and $Re_p$ ranged from 328-948. The geometry variations associated with plane PL-2 through PL-4 and their role on transition are also included.

\begin{table}
\begin{center}
\caption{Uncertainty estimates for RIM porous media.}
\label{tab:uncertainty}

\begin{tabular}[c]{@{}p{0.75\linewidth}@{}p{0.17\linewidth}@{}p{0.08\linewidth}@{}}
\hline\hline
\textbf{Uncertainty Source}&\textbf{$\varDelta$x [px]}&\textbf{$\varDelta$y [px]}\\
\hline
\text{Max displacement}&  4.45&  8.69\\
\text{In-plane loss of image pairs (bias)}&  0.091&  0.091\\
\text{Finite number of image pairs (random)}& $\rm 0.037$ &$0.037$ \\
\text{Refractive index mismatch error}& $0.088$&$0.088$\\
\text{Magnification}& $0.07$&$0.012$\\
\text{Displacement gradients}&$0.011$ &$0.056$ \\
\text{Out-of-plane motion (perspective)} & $0.097$ & $0.099$\\
\text{Total error}& $0.39$&0.4\\
\text{Total error}~ [$\%$ of max displacement]& $8.8$&$4.6$\\
\hline\hline

\end{tabular}
\end{center}
\end{table}
 
 \subsection{Uncertainty estimation}\label{subsec:uncertainty}
 Comprehensive RIM-PIV uncertainty estimation is estimated based on six major sources of error addressed by Patil and Liburdy \cite{patil2012optical,patil2013flow,patil2013turbulent}, Adrian and Westerweel\cite{adrian2011particle}, Raffel et al. \cite{raffel2018particle}, Northrup et al.\cite{northrup1993direct}, Sciacchitano et al. \cite{sciacchitano2016piv}. Uncertainty analysis was performed on all 3200 instantaneous PIV data collected for all six planes for $Re_p$ ranging from 100-948.  Uncertainty sources are listed in \hyperref[tab:uncertainty]{Table~\ref*{tab:uncertainty}}. The light refraction uncertainties from the glass bead surfaces and test-section wall were minimized by using the optimal focal length and polarized laser sheet. The total uncertainty is quantified relative to the maximum displacement derived from the PL-6 velocity field. The maximum displacements in this experiment for the highest pore Reynolds number in the longitudinal or stream-wise direction (Y) is 8.69 pixels, while the lateral direction (X) experiences a maximum measured displacement of 4.45 pixels. The total error in the stream-wise direction (longitudinal) as a fraction of total y-displacement is 4.6$\%$, while it is 8.8$\%$  of maximum x-displacement in the cross-stream (lateral) direction.

The uncertainty associated with refractive-index mismatch is characterized by the method described in Patil and Liburdy \cite{patil2013flow}, and the magnification error follows that described by Patil and Liburdy \cite{patil2012optical}, while other sources were found based using typical PIV error estimations \cite{sciacchitano2016piv,adrian2011particle,raffel2018particle}. For estimating the in-plane loss of image pairs (bias error) as well as finite number of seed images in each interrogation window (random error), image shifting was used from 0 to 8 pixels with one pixel increments. The seed image displacement at the subpixel increments was applied based on interpolation using the nearest neighborhood. The total out-of-plane motion of particles were estimated to be 24$\%$ of the total error using the method given in (Ref.\onlinecite{prasad1993stereoscopic}). The velocity gradient error was determining by using central difference scheme \cite{westerweel2008velocity,adrian2011particle}. At the highest Reynolds number ($Re_p = 948$), the velocity gradients uncertainty based on the propagation of random error is estimated to be 3.67 $s^{-1}$ which is 1.98$\%$ of max.

\section{Results and discussion}\label{sec:results}
A detailed analysis on the scales associated with vortical structure characteristics is presented in this section. In summary, this section represents (i) the mean flow analysis with emphasis on flow structures  (\hyperref[subsec:mean]{Section~\ref*{subsec:mean}}), (ii) global scale estimation of vortical flow structures in different $Re_p$ (\hyperref[subsec:global]{Section~\ref*{subsec:global}}), and (iii) local (pore-scale) scale estimation of vortical structures for the range of $Re_{\langle p\rangle}$ (\hyperref[subsec:local]{Section~\ref*{subsec:local}}).

\begin{figure*}
 \centering
 \includegraphics[width=0.97\textwidth]{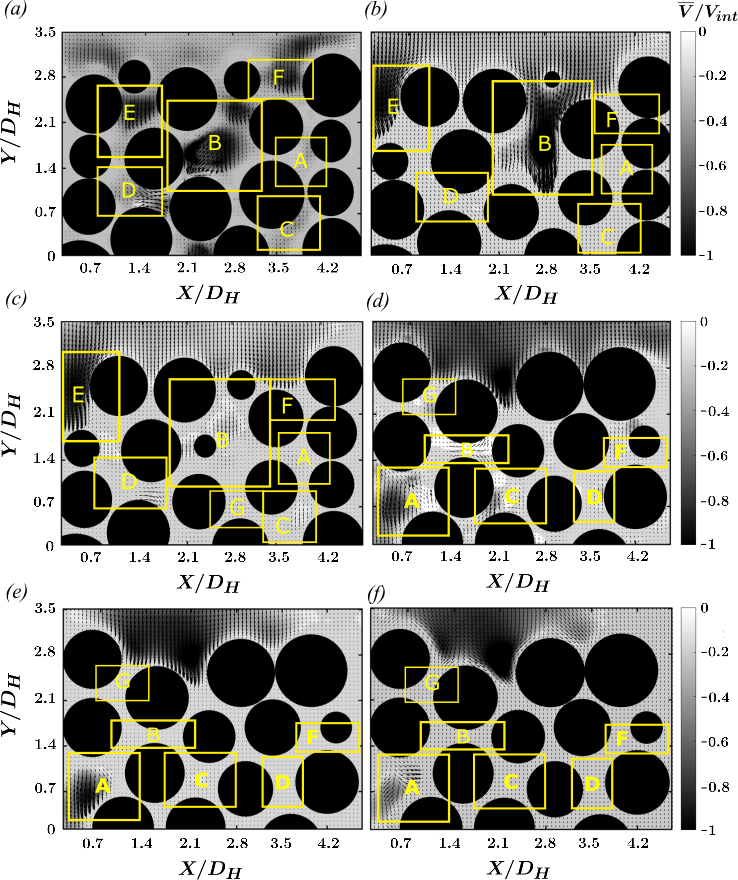}
 \caption{Contours of non-dimensional longitudinal velocity $\overline{V}/V_{int}$ overlayed with time-averaged normalized vector-map configuration of four different planes for Reynolds numbers ranging from 100 to 948; (a) Plane PL-1 at $Re_{p}=100$, (b) Plane PL-3 at $Re_{p}=272$, (c) Plane PL-4 at $Re_{p}=270$, (d) Plane PL-6 at $Re_{p}=528$, (e) Plane PL-6 at $Re_{p}=736$, and (f) Plane PL-6 at $Re_{p}=948$.}
\label{fig:time-avg}
 \end{figure*}

\subsection{Mean Flow characteristics } \label{subsec:mean}

Results of the mean velocity distribution are first presented to better understand the global effects of $Re_p$ within the entire bed. Referring to \hyperref[fig:8]{Figure~\ref*{fig:8}}, four planes of PL-1, PL-3, PL-4, and PL-6 were selected to study the variation of mean flow. The effect of Reynolds number ($Re_p$) and location of the plane relative to the bed was considered in studying the mean flow variations. The chosen planes were located in the middle of the bed (PL-1, PL-3, PL-4) and near the wall region (PL-6). 

 Shown in \hyperref[fig:time-avg]{Figure~\ref*{fig:time-avg}}, the variation of random geometrical arrangement of pores is a major contributor to have different distributions of mean velocity within the entire bed. Contours of time-averaged two-component velocity magnitude ($\overline{V}$) normalized by the interstitial velocity, $V_{int}$ are illustrated in \hyperref[fig:time-avg]{Figure~\ref*{fig:time-avg}}. The contours are overlayed with the normalized vector field demonstrating the local direction of the velocity at each specific pore. In this graph, local pores are indicated with rectangular regions representative of the area boundaries for selected regions to investigate the local flow structures. Selected pore-regions exhibit different flow characteristics in a heterogeneous pattern. The heterogeneity of the flow due to geometry has been observed by Johns et al. (2000) \cite{johns2000local}, and Sederman et al. (1998) \cite{sederman1998structure} in the axial versus transverse directions, which is also discussed by Ogawa et al. (2000) \cite{ogawa2000correlation} analogous to the axial variation of velocity magnitude shown in \hyperref[fig:time-avg]{Figure~\ref*{fig:time-avg}}. Furthermore, changes to the mean flow Reynolds number ($Re_p$) distributes the flow differently in each pore as shown in \hyperref[fig:time-avg]{Figure~\ref*{fig:time-avg}}. In addition to inertial effects, the viscous shear effects also contributes to the mean flow variations as it has been discussed by Hill and Koch (2001) \cite{hill2001moderate} in a random packing structure. The pore-scale analysis discussed later in \hyperref[subsec:local]{Section~\ref*{subsec:local}} is based on the selected pores indicated in \hyperref[fig:time-avg]{Figure~\ref*{fig:time-avg}}.

The mean flow with a nearly uniform velocity profile enters the porous bed and accelerates or decelerates within the constriction regions in the interstices and becomes non-uniform depending on the pore-scale geometry causing the changes in pore-scale pressure, and velocity distribution. Therefore, a combination of different types of flow within these pores is created at different Reynolds numbers. The distribution of velocity is more homogeneous for $Re_p$ larger than 300, as shown in \hyperref[fig:time-avg]{Figure~\ref*{fig:time-avg}}.

\hyperref[fig:time-avg]{Figure~\ref*{fig:time-avg} (a)} shows representative pores alphabetically ordered from A to F at a low Reynolds number of 100 (Plane PL-1). There are regions where flow behaves similar to a recirculation zone such as pore B, or a nozzle-like flow such as pore D. Also, in Figures \hyperref[fig:time-avg]{\ref*{fig:time-avg} (a)} and  \hyperref[fig:time-avg]{\ref*{fig:time-avg} (b)}, there are wake-like regions, as in pore B, where flow experiences a higher velocity. A wake extends behind the sphere in these pores on an order of 1-1.5$D_H$ and the flow accelerates downstream. However, in \hyperref[fig:time-avg]{\ref*{fig:time-avg} (c)}, pore B represents a fairly uniform velocity distribution, which is not influenced by the wake region as opposed to planes PL-1 and PL-3 in Figures \hyperref[fig:time-avg]{\ref*{fig:time-avg} (a)}, and \hyperref[fig:time-avg]{\ref*{fig:time-avg} (b)}, respectively. Here, pore B is a three-dimensional void space that extends between PL-1 to PL-4, where the flow topology not only varies longitudinally but also the velocity distribution is affected in the transverse direction (z) as well. Figures \hyperref[fig:time-avg-pore]{\ref*{fig:time-avg-pore} (a)} to \hyperref[fig:time-avg-pore]{\ref*{fig:time-avg-pore} (c)} show the pore-scale contours of non-dimensional velocity magnitudes $\overline{V}/\langle \overline{V_{\langle P\rangle}}\rangle$ in pore B for PL-1, PL-3, and PL-4 mentioned above. The wake regions are magnified to show the pore-scale velocity distribution in pore B using three planes.

The PL-6 located nearer to the side walls, being 1.5 $D_B$ away from the wall. This plane has been shown in Figures \hyperref[fig:time-avg]{\ref*{fig:time-avg} (d)} to \hyperref[fig:time-avg]{\ref*{fig:time-avg} (f)} for three values of $Re_p$ of 528, 736, and 948. By increasing $Re_p$ the interestitial velocity also increases, hence the ratio of $\overline{V}/V_{int}$ becomes more homogeneous. There are two reasons for this: one, $Re_p$ has been increased such that the steady wake-like structures, or recirculation zones, are diminished by the mean flow inertia, and two, the porosity of PL-6 is smaller than the other planes, hence large flow patterns are damped due to higher geometrical confinements. Similar flow distribution effects has been reported by Hill et al. (2001)\cite{hill2001moderate}. In PL-6, there are other pores that show other flow patterns such as a recirculation region shown in pore B in \hyperref[fig:time-avg]{Figure~\ref*{fig:time-avg} (d)}, where the flow occurs in a much smaller area, hence the recirculation is approximately 1/3$D_B$. Figures \hyperref[fig:time-avg-pore]{\ref*{fig:time-avg-pore} (d)} to \hyperref[fig:time-avg-pore]{\ref*{fig:time-avg-pore} (e)} also represent an expanded view of pore A in PL-6. The pore-scale contours of $\overline{V}/\langle \overline{V_{\langle P\rangle}}\rangle$ shows more homogeneity as $Re_p$ increases.

The local (pore-scale) length scale of the pores governs the flow structure sizes which is characterized by the local pore Reynolds number. These wake-like separated flows behind spheres have also been observed by Patil and Liburdy (2013) \cite{patil2013turbulent} at a pore Reynolds number of 418 with almost similar porosity, where the wake extends about 0.24$D_H$. On the other hand, Hill and Koch (2001) \cite{hill2001moderate} simulated the wake structures in a dilute random array of spheres with higher porosity of 0.7 to 0.9. They hypothesized that the wake extends no further than an $O(\rm D_B/2(1-\phi))$ in moderate $Re_p\approx300$ with $\phi\approx0.75-0.9$. This is calculated to be in the order of $D_H$ in the present study, which is consistent with what is observed at $Re_p$ less than 300 in some pores. These wakes have also been observed in regular packing in numerical simulations. Chu et al. (2018) \cite{chu2018flow} reported such wake-like structures at a higher Reynolds number ($Re_p$ = 1000) with porosity of 0.75 in a regular array of square cylinders. He and Apte \cite{he2019characteristics} investigated weak to strong wakes in a Face-Centred Cubic (FCC) array at pore Reynolds numbers 300, 500, and 1000 with low porosity of 0.26. They state that the wake-structure shows an organized (coherent) pattern which is due to the geometrical symmetry of the porous bed. However, as it is seen in \hyperref[fig:time-avg]{Figure~\ref*{fig:time-avg}}, the flow structures are artifacts of complex combination of flow inertia, and randomness of geometrical constrictions, as well as flow geometry heterogeneity. The porosity is another factor in the size of flow structures in porous media; larger wavelengths occur behind spheres in a high porosity bed.
\begin{figure*}
 \centering
 \includegraphics[width=0.97\textwidth]{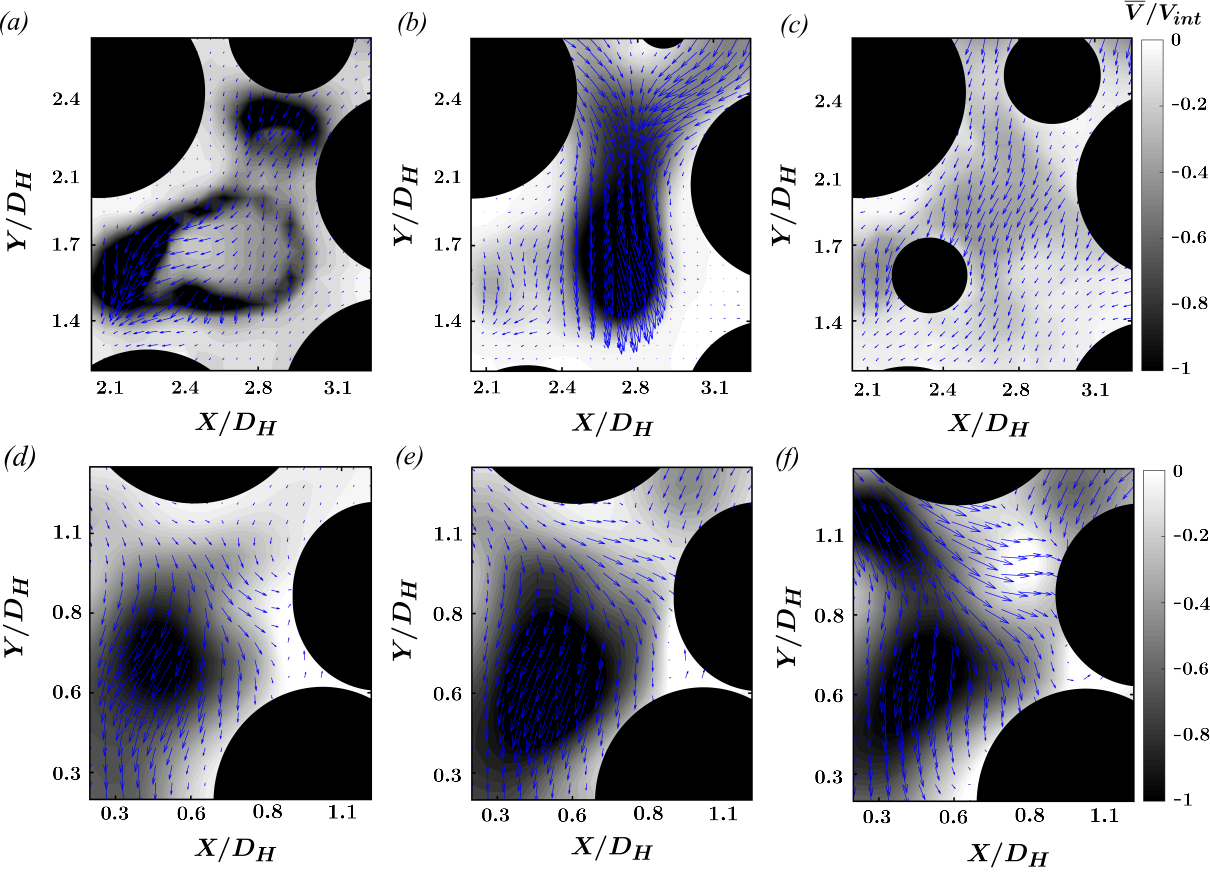}
 \caption{Pore-scale contours of non-dimensional longitudinal velocity $\overline{V}/\langle \overline{V_{\langle P\rangle}}\rangle$ overlayed with the time-averaged normalized vector-map of four planes for Reynolds numbers ranging from 100 to 948; (a) pore B in plane PL-1 at $Re_{p}=100$, (b) pore B in plane PL-3 at $Re_{p}=272$, (c) pore B in plane PL-4 at $Re_{p}=270$, (d) pore A in plane PL-6 at $Re_{p}=528$, (e) pore A in plane PL-6 at $Re_{p}=736$, and (f) pore A in plane PL-6 at $Re_{p}=948$.}
\label{fig:time-avg-pore}
 \end{figure*}

High or low velocity regions in different pores are also the result of the tortuous three-dimensional flow in porous media. This feature is observed in \hyperref[fig:time-avg-pore]{Figure~\ref*{fig:time-avg-pore}}. Pore B (for PL-1, PL-3, and PL-4) shown in Figures \hyperref[fig:time-avg-pore]{\ref*{fig:time-avg-pore} (a)} to \hyperref[fig:time-avg-pore]{\ref*{fig:time-avg-pore} (c)}, and Pore A for PL-6 shown in Figures \hyperref[fig:time-avg-pore]{\ref*{fig:time-avg-pore} (d)} to \hyperref[fig:time-avg-pore]{\ref*{fig:time-avg-pore} (f)} demonstrate high velocity pores that are results of the pore-scale out-of-plane motion of the flow in some specific pores within the bed. Moreover, the effect of spheres located outside of the plane of visualization sheds into the plane and a complex flow patterns is created with intermittent regions of high velocity as seen in pore B in \hyperref[fig:time-avg-pore]{Figure~\ref*{fig:time-avg-pore}(a)}. As the plane of PIV moves closer to the wall, the normalized velocity in the porous bed exhibits a more uniform distribution in the entire plane. 

\subsection{Global scale of vortical flow structures} \label{subsec:global} 

In this section, scales of the vortical structures for transitional flow are analyzed for the flow within the entire bed. Time-resolved PIV technique used in this study follows the Eulerian method of capturing velocity vectors instantaneously. Therefore, Eulerian vortex identification methods are used to detect the vortical structures. The swirl strength, known as point-wise (local) detection method of vortical structures ($\lambda_{ci}$, enhanced swirl strength ($\lambda_{cr}/\lambda_{ci}$) versus multiple-point (non-local) detection method of vortex core ($\varGamma_{2}$) are used \cite{epps2017review,zhou1999mechanisms,kolavr2007vortex}. Unique features of the evolution of vortical structures based on the integral length scale obtained from autocorrelation and also vortical size estimates derived from spatial distribution of vortex strength ($\lambda_{ci}$), vortex boundary ($\varGamma_{2}$), and size of spiral vortices ($\lambda_{cr}/\lambda_{ci}$). Also, strength of the vortex ($\lambda_{ci}$) as a vortex time-scale is compared with integral time-scales of the flow. Lastly, the number density of the vortical structures is found according to the number identified vortical structures based on $\lambda_{ci}$ criterion and the $\varGamma_{2}$ technique proposed by Michard et al. (1997) \cite{michard1997identification} and Graftieaux et al. (2001)\cite{graftieaux2001combining}. 

The critical point analysis is based on the point-wise (local) linear approximation of the kinematics of local fluid motion at each point. Therefore, the local instantaneous velocity $\boldsymbol{V}$ around a point following the work proposed by Chong et al. (1990) \cite{chong1990general} denoted by a position vector $\textit{r}$ can be expressed, to linear order, as \cite{zhong1998extracting,chen2015comparison,zhou1999mechanisms}:
 
 \begin{equation}
\boldsymbol{V}(\textit{r}+\delta \textit{r}) = \boldsymbol{V}(\textit{r}) + D_{ij} \ \delta \textit{r} + \textit{O}(\|\delta r\|^2)
\label{eq:u}  
\end{equation}

\noindent where, $D_{ij} = \nabla \boldsymbol{V}$ is the velocity gradient tensor with the characteristic equation of $D_{ij}$ is given by:

\begin{equation}
\lambda^2 + P \lambda + R = 0
\label{eq:characteristic}  
\end{equation}

\noindent where \textit{P} and \textit{R} are the two invariants of $D_{ij}$, which are defined as $P=\frac{\partial u}{\partial x} + \frac{\partial v}{\partial y}$, and $R=\frac{\partial u \partial v}{\partial x \partial y} + \frac{\partial u \partial v}{\partial y\partial x}$ where $u$ and $v$ are the $x$, and $y$ components of the velocity $\boldsymbol{V}$. By computing the imaginary component of the complex conjugate of the eigenvalue of the velocity gradient tensor, the field variable $\lambda_{ci}$ (local swirling strength) is determined for the flow\cite{chong1990general,zhou1999mechanisms,chen2015comparison}. This criterion is derived using a requirement, where the discriminant of characteristic equation becomes negative, then in this case, the discriminant has a conjugate pair of complex eigenvalues ($\lambda_{cr}\pm i\lambda_{ci}$).

 \begin{figure*}
 \centering
 \hspace{0.09\textwidth}
 \includegraphics[width=1.0\textwidth]{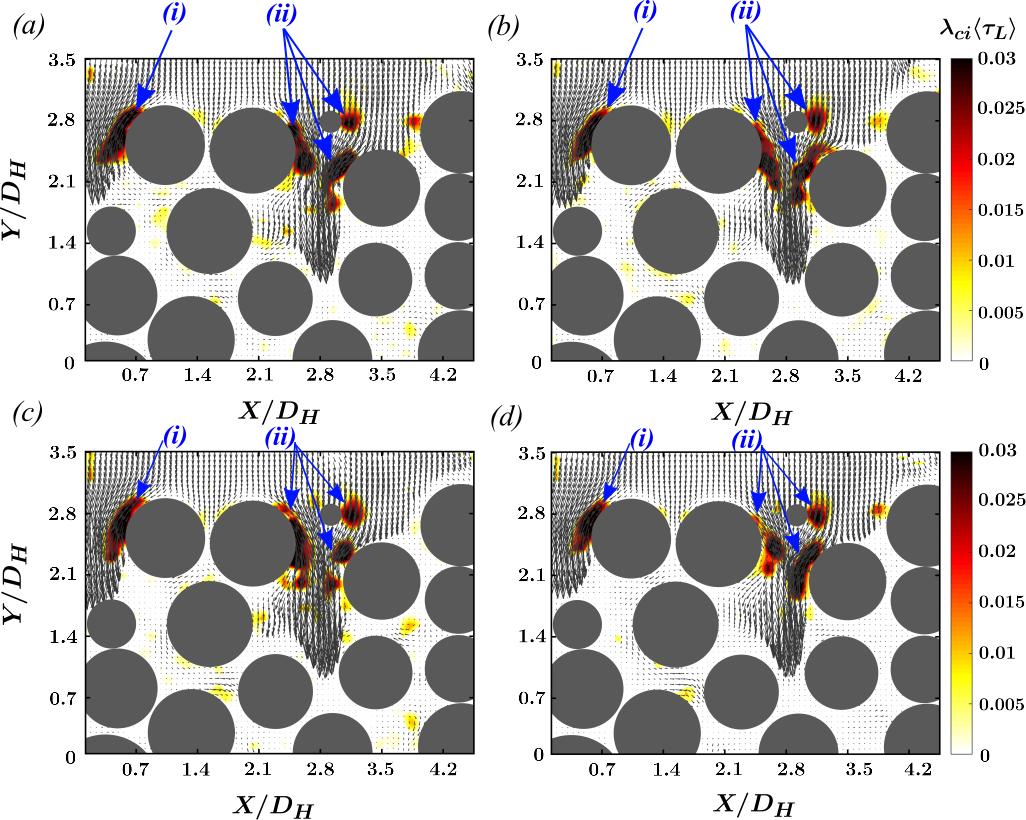}
 \caption{(a)-(d) Contours of the instantaneous swirl strength of detected vortical structures normalized by spatially averaged longitudinal Eulerian integral time scale ($\langle \tau_L \rangle$) overlayed on the LES filtered velocity with Gaussian kernel of $3\times3$ plotted for $Re_{p} = 270$ at four consecutive time intervals with the time-steps of $16.2\ ms$.}
\label{fig:Lambda-time}
 \end{figure*}
 
 The condition through which vortical structures evolve with changing $Re_p$ differ due to topology of the pores, and local pore Reynolds number ($Re_{\langle p\rangle}$). To better observe the scale of flow structures, Large Eddy Scale (LES) filtering has been applied to the instantaneous velocity to separate the large scale flow structures from small scale vortices. A series of top-hat and Gaussian kernels have been convolved with each instantaneous velocity field \cite{adrian2000analysis}. A selected optimal kernel size for filtering based on $Re_p$ is performed. In laminar flow ($Re_p \le 300$), a Gaussian filter with the kernel size of 5$\times$5 ($0.31D_H$) extracts the larger eddies from the unfiltered velocity, while for inertial and turbulent Reynolds numbers ($Re_p > 300$), a Gaussian filter with a size of 3$\times$3 ($0.18D_H$) is used. In \hyperref[fig:Lambda]{Figure~\ref{fig:Lambda-time}}, the contours of instantaneous $\lambda_{ci}$ are overlayed with LES filtered velocity using the Gaussian kernel of $3\times 3$.

 In \hyperref[fig:Lambda-time]{Figure~\ref{fig:Lambda-time}}, the global averaged Eulerian integral time scale for plane PL-4 ($Re_p = 270$) is evaluated to be $\langle \tau_{L}\rangle = 17.6~ms$ and the time interval between images is approximately $\langle \tau_{L}\rangle$. $\langle\tau_{L}\rangle$ is evaluated using the temporal autocorrelation of the longitudinal fluctuating velocity. The average vortex strength in this plane is 0.62 $s^{-1}$, showing that the frequency of the detected structures is much smaller than the sampling frequency for the data set. The dark regions of high $\lambda_{ci}$ near the surface of spheres shown by regions (i) and (ii) in \hyperref[fig:Lambda-time]{Figure~\ref{fig:Lambda-time}} are positioned similarly with almost identical cores in four consecutive time steps. These regions are consequences of high velocity gradients near the surface of spheres at a moderate $Re_p$ of approximately 300. They maintained their structures approximately within the time period shown in \hyperref[fig:Lambda]{Figure~\ref{fig:Lambda-time}}, hence they are stretched but not broken up and carried with the mean flow inertia due to viscous effects near surfaces. The sequence shows distinct distortion over time as it is stretched by the mean flow. Speculated from the event shown in \hyperref[fig:Lambda]{Figure~\ref{fig:Lambda-time}}, the vorticity develops nearly at distinct locations (due to highly tortuous bed structure) but with higher concentrations near the surface of spheres. These vortices are mostly stretched relative to the mean flow directions similar to the same behavior of vortices simulated in $Re_p = 300$ for a regular packed bed \cite{he2019characteristics}.
 
 \begin{figure*}
 \centering
 \hspace{0.09\textwidth}
 \includegraphics[width=1.0\textwidth]{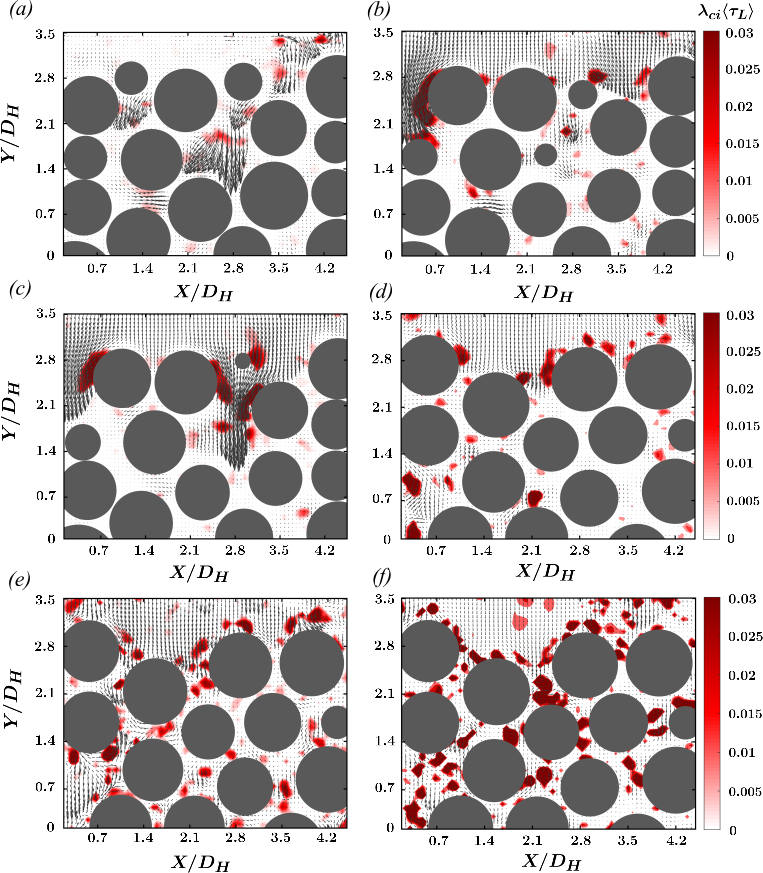}
 \caption{Contours of the instantaneous swirl strength of detected vortical structures normalized by the longitudinal Eulerian integral time scale ($\lambda_{ci}\tau_L$) overlayed on the LES filtered velocity with Gaussian kernel of $3\times3$ plotted for a time step of $T = 20\Delta\tau$ for all six cases of \hyperref[fig:time-avg]{Figure~\ref*{fig:time-avg}}. (a) Plane PL-1 at $Re_{p}=100$, (b) Plane PL-3 at $Re_{p}=272$, (c) Plane PL-4 at $Re_{p}=270$, (d) Plane PL-6 at $Re_{p}=528$, (e) Plane PL-6 at $Re_{p}=736$, and (f) Plane PL-6 at $Re_{p}=948$.}
\label{fig:Lambda}
 \end{figure*}

The results of evaluating the swirl strength, $\lambda_{ci}$, in six different planes of interest are given in \hyperref[fig:Lambda]{Figure~\ref{fig:Lambda}}. The value of $\lambda_{ci}$ is normalized by the inverse of the bed averaged longitudinal integral time scale of the flow corresponding to each $Re_p$, $\langle\tau_{L}\rangle$. This set of figure shows the instantaneous vortical structures, and the location of vortical structures as they change while moving through the bed. However, for the lower Reynolds numbers (i.e $Re_p\le 300$), many of the pore-scale structures are invariant in size and location; the uniform size of these steady coherent vortices develop near the surfaces as a consequence of shear dominated flows. Smaller rotational structures are weaker and are advected downstream. With increasing Reynolds number, these structures stretch, and break into smaller eddies as shown in Figures\hyperref[fig:Lambda]{~\ref{fig:Lambda} (a)} to \hyperref[fig:Lambda]{\ref{fig:Lambda} (f)}.

Analogous to high velocity gradient regions with large swirl strength ($\lambda_{ci}$) shown in \hyperref[fig:Lambda]{Figure~\ref{fig:Lambda-time}}, the lower $Re_p$ conditions shown in \hyperref[fig:Lambda]{Figure~\ref{fig:Lambda}} (for cases with $Re_p$<300), contain vortical structures that are shear dominant as they tend to be generated near the surfaces of spheres in high velocity gradient regions. In higher $Re_p$ flows, these structures are advected downstream, where they are broken up to smaller eddies. To further reveal the regions with high velocity gradients, the instantaneous vorticity for pore B (which has a local high velocity) in PL-4 ($Re_p = 270$) is discussed in \hyperref[fig:vort-time]{Figure~\ref{fig:vort-time}}. The time history of four consecutive maps of non-dimensional out-of plane vorticity is illustrated for $Re_p = 270$ (PL-4). The shear-dominant high vorticity regions are larger and have higher swirl strength (frequency) similar to regions (i), and (ii) in \hyperref[fig:Lambda]{Figure~\ref{fig:Lambda-time}}. The generation of shear dominant vorticity near the surfaces and their elongation in the direction of mean flow have also been observed in other random packed beds such as (Ref. \onlinecite{finn2012characteristics}) at smaller $Re_p = 150$ with $\phi = 0.47$, as well as (Ref. \onlinecite{nguyen2018time}) at larger $Re_p = 340$ with porosity of 0.42. Comparing \hyperref[fig:Lambda]{Figure~\ref{fig:Lambda}} with \hyperref[fig:vort-time]{Figure~\ref{fig:vort-time}} illustrates the large areas of vorticity, where the $\lambda_{ci}$ criterion identifies the regions with intense vorticity and weakly detects the regions where the vorticity is in balance with strain rate \cite{cantero2008turbulent}. 

\begin{figure*}
 \centering
 \includegraphics[width=0.7\textwidth]{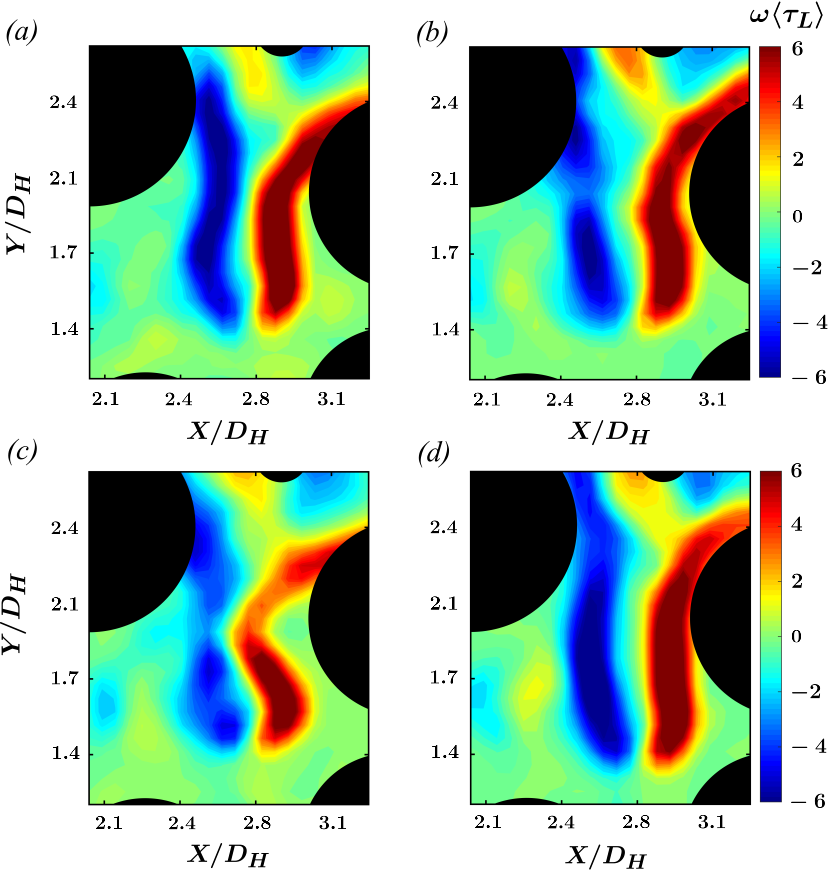}
 \caption{(a)-(d) Sequences of instantaneous out-of-plane vorticity contours of detected vortical structures normalized by spatially averaged inverse longitudinal Eulerian integral time scale ($\langle \tau_L \rangle$) plotted for $Re_{p} = 270$ (PL-4) at four consecutive time steps with interval of 16.2 ms.}
\label{fig:vort-time}
\end{figure*}

\begin{figure*}
 \centering
 \includegraphics[width=0.98\textwidth]{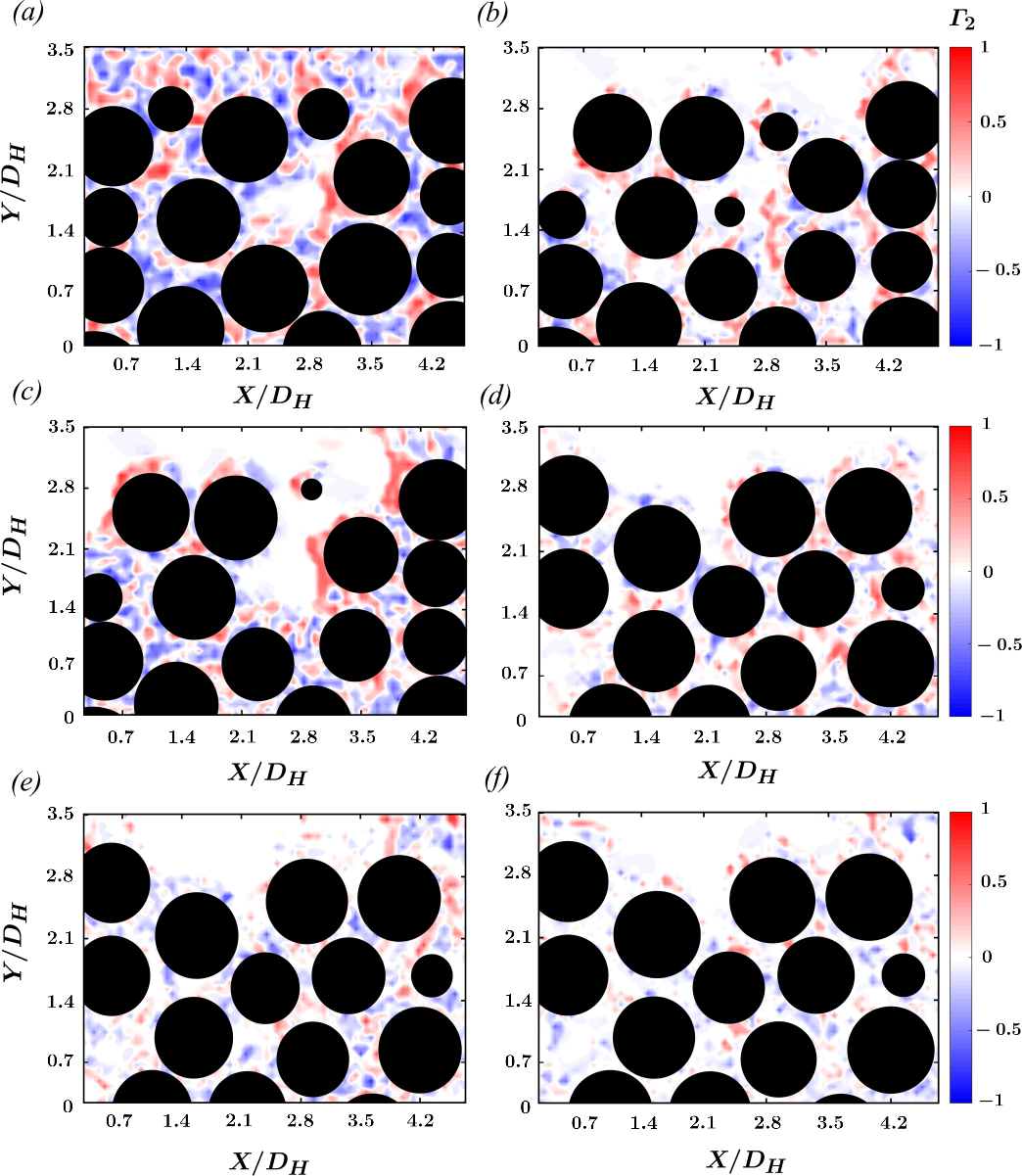}
 \caption{Instantaneous vortex core structures identified by $\Gamma_{2}$ scheme with a $3\times3$ kernel size corresponding to 0.12$D_B$. (a) Plane PL-1 at $Re_{p}=100$, (b) Plane PL-3 at $Re_{p}=272$, (c) Plane PL-4 at $Re_{p}=270$, (d) Plane PL-6 at $Re_{p}=528$, (e) Plane PL-6 at $Re_{p}=736$, and (f) Plane PL-6 at $Re_{p}=948$. Computed values are for the whole porous bed.}
\label{fig:Gamma2}
\end{figure*}

Distinguishing between the rotational and shear dominant flow structures was employed by implementing $\varGamma_2$ function introduced by Graftieaux et al. (2001) \cite{graftieaux2001combining}, which is derived using velocity vector field directly. Thus, the $\varGamma_2$ criterion utilizes pseudo-circulation \cite{epps2017review,ford2013lift} to identify vortex cores defined as: 
 
\begin{equation}
\rm\varGamma_2(X; \textit{R}) = \frac{1}{\pazocal{S}}\oint\limits_{X^\prime\in \pazocal{S} }\sin \theta \ \textit{d} X^\prime
\label{eq:gamma2}  
\end{equation}

\noindent where $\pazocal{S}$ is a disc surrounding point $\rm X$, $\rm X^\prime$ is any point surrounding point $\rm X$ in 2D PIV flow field within each $3\times 3$ subregion, $V$ is the velocity vector, $\sin \theta = |\delta X \times \delta V|/|\delta X||\delta V|$, $\delta X = X^\prime - X$, and $\delta V = V(X^\prime)-\overline{V(X)}$, and $\overline{V(X)}$ is the average velocity evaluated in the subregion. The vortex core was then computed for eight vectors surrounding the point of interest in the flow field via $\varGamma_2=\frac{1}{8}\sum_{i=1}^{8} \sin \theta_i$. As a robust method that depends spatially on the velocities in a small neighborhood (non-local \cite{epps2017review}), the subregion sweeps all over the vector field to detect vortex cores.

The vortical structures detected by the instantaneous $\varGamma_2$ criterion  is given in \hyperref[fig:Gamma2]{Figure~\ref{fig:Gamma2}}. The contours are  color-coded based on the degree of fully rotational flow in the vortex center ($\lvert\varGamma_2\rvert = 1$) versus the fully shear flow at the boundary of vortex ($\lvert\varGamma_2\rvert = 0$). Therefore, a complete vortex zone is closer to 1, with positive sign representing counter-clockwise, while negative values describe clockwise flow \cite{epps2017review,morgan2009vortex}. The shear dominant structures are identified when $\varGamma_2$ is less than $2/\pi$, while the rotational structures overcome the shear structures when $\varGamma_2\ge 2/\pi$ \cite{graftieaux2001combining}. 

It is seen that in general, the size of the identified flow structures decreases with increasing $Re_p$. At $Re_p = 100$ (\hyperref[fig:Gamma2]{Figure~\ref{fig:Gamma2}(a)}), the flow is essentially laminar and flow structures are larger and dominated by shear, while the results for $Re_p = 948$ (\hyperref[fig:Gamma2]{Figure~\ref{fig:Gamma2}(f)}) displays relatively small distributed turbulent flow structures. 

\begin{figure*}
 \centering
 \includegraphics[width=1\textwidth]{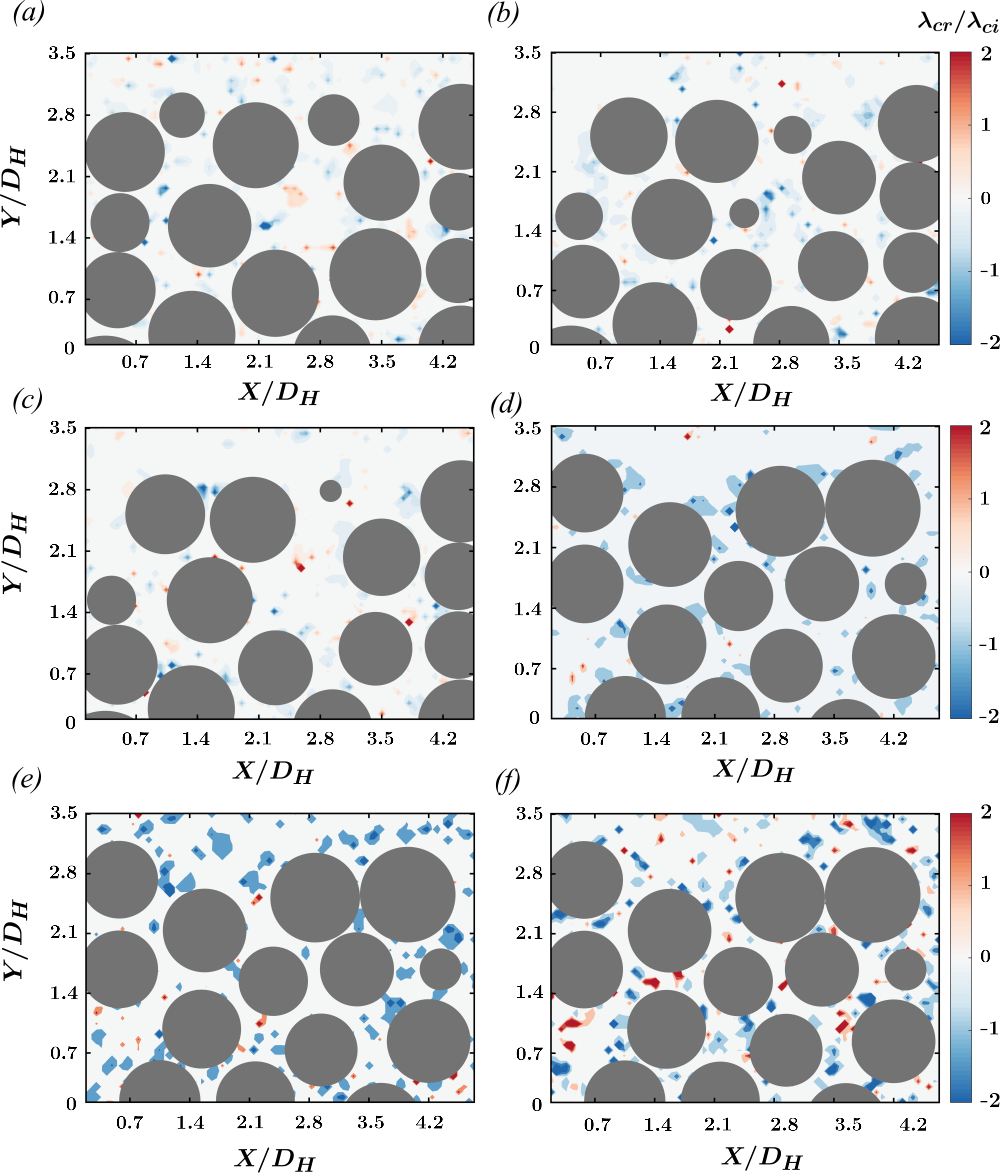}
 \caption{Instantaneous $\lambda_{cr}/\lambda_{ci}$ criterion indicating vortical structures with spiral motions passing through the plane of visualization at (a) Plane PL-1 at $Re_{p}=100$, (b) Plane PL-3 at $Re_{p}=272$, (c) Plane PL-4 at $Re_{p}=270$, (d) Plane PL-6 at $Re_{p}=528$, (e) Plane PL-6 at $Re_{p}=736$, and (f) Plane PL-6 at $Re_{p}=948$.}
\label{fig:lambda_cr}
\end{figure*}

The $\lambda_{ci}$ criterion for vortex strength identification is Galilean invariant, which locally translates with the frame of reference\cite{chakraborty2005relationships}. The non-local identification of swirling structures \cite{kolavr2007vortex} and their level of compactness or orbital shape is distinguished using $\lambda_{cr}/\lambda_{ci}$ (Chakraborty et al., 2005\cite{chakraborty2005relationships}). Using this method, the ability to detect the size of structures that are prone to elongate out of the plane of visualization is possible. Results are shown in \hyperref[fig:lambda_cr]{Figure~\ref{fig:lambda_cr}} for the size and number density of spiral vortices. It is obvious that the size, strength, and number density increases with $Re_p$. The number of spiral structures increases as they advect downstream in the y-direction due to the compactness of the pores, however, the size of the vortices are reduced. This is also observed based on $\lambda_{ci}$ (\hyperref[fig:Lambda]{Figure~\ref{fig:Lambda}}), as well as $\varGamma_{2}$ (\hyperref[fig:Gamma2]{Figure~\ref{fig:Gamma2}}). As stated by Chakraborty et al. (2005) \cite{chakraborty2005relationships}, the presence of viscosity in real fluids results in the continuity of kinematics, which gives rise to the proximity of the estimates of vortical structures from $\lambda_{ci}$ as a local (pointwise, applied point by point) and the non-local method of $\varGamma_{2}$. 

With increasing $Re_p$ shown in \hyperref[fig:lambda_cr]{Figure~\ref{fig:lambda_cr}}, the size of detected vortices in the threshold of $-2\le \lambda_{cr}/\lambda_{ci}\le 2$ increases slightly in compliance with the three-dimensional tortuous effects of the porous media; thereby contributing to more spiral effects in vortical structures. The positive values of $\lambda_{cr}/\lambda_{ci}$ corresponds to outward spiraling motion of vortex, while  negative values address the inward spirals.  The areas where no vortex is detected based on $\lambda_{cr}/\lambda_{ci}$ corresponds to regions where there is no swirling in the flow since $\lambda_{ci}=0$. Thereby, the eigenvalue of $\nabla V$ does not have an imaginary pair in the regions with white color in \hyperref[fig:lambda_cr]{Figure~\ref{fig:lambda_cr}}. The swirling spiral structures with intense out of plane motions are observed for $Re_p$ greater than 500. 

\begin{figure*}
 \centering
 \includegraphics[width=1\textwidth]{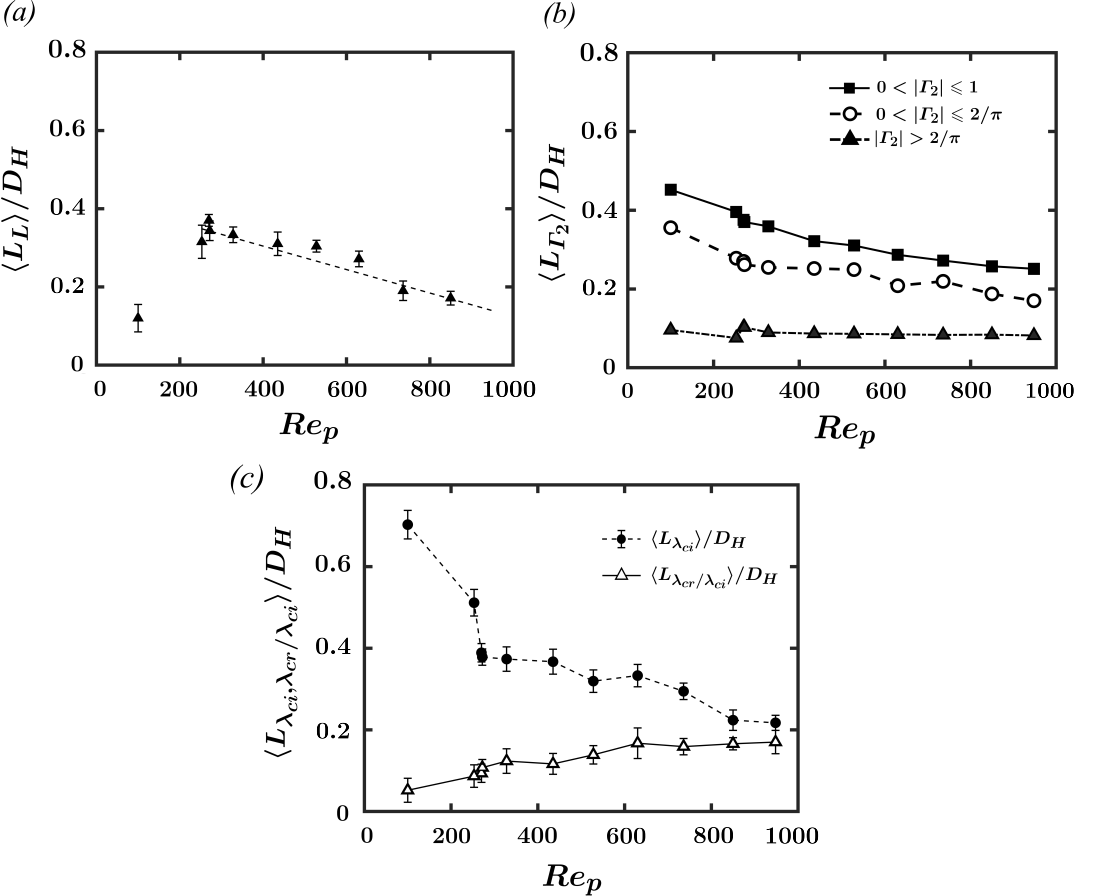}
 \caption{Vortex size in the entire porous bed: (a) the variations of vortex size computed based on longitudinal integral length scale ($\langle L_L\rangle$) normalized by $ D_H$, (b) vortex size computed based on vortex boundary scheme ($\Gamma_{2}$) for the entire vortex core size ($\lvert \Gamma_{2} \rvert\leq 1$), shear-dominant votices ($\lvert \Gamma_{2} \rvert\leq 2/\pi$), and rotation-dominant vortices ($\lvert \Gamma_{2} \rvert\ge 2/\pi$), and (c) vortex size based on $\lambda_{ci}$ ($L_{\lambda_{ci}}$), as well as $\lambda_{cr}/\lambda_{ci}$ criteria. Computed values are scaled by hydraulic diameter $D_H$ for the whole porous bed for the cases of Reynolds numbers given in \hyperref[tab:cases]{Table~\ref*{tab:cases}} .}
\label{fig:L_ci}
 \end{figure*}

The size of vortical structures is detected with the normalized Eulerian longitudinal integral length scale averaged over the entire porous bed for all six planes ($\langle L_{L}\rangle/D_H$) and is shown in \hyperref[fig:L_ci]{Figure~\ref{fig:L_ci} (a)} for the entire range of $Re_p$. The trend shows a decay of size with increasing $Re_p$ beyond 300. However, at $Re_p = 100$ there is a reduced length scale ($\langle L_{L}\rangle/D_H$ = 0.12) since the has not transitioned to turbulence. 

The $\varGamma_2$ vortex boundary method provides a means to predict the size of each vortex using these conditions: (i) $0<\varGamma_2\le 1$, (ii) $0<\varGamma_2\le 2/\pi$, and  (iii) $\varGamma_2\ge 2/\pi$. The first describes all vortical structures (shear and rotational), while the second refers to shear dominated vortices, and the latter refers to the rotational structures with the condition of no vortex described by $\varGamma_2 = 0$.  Similarly, the $\lambda_{ci}$ method also detects an area associated with each vortex based on a given threshold value. The size of each vortex was computed through image processing of the binarized image of detected vortices using both the $\varGamma_2$ vortex boundary method and swirl-strength criteria  ($\lambda_{ci}$). Assuming a circular area for vortical structures, the diameter was found as a function of $Re_p$ which asymptotically converges to the value of 0.25$\rm D_H$ as shown in \hyperref[fig:L_ci]{Figure~\ref{fig:L_ci} (b)} along with the integral length scale. As shown, increasing $Re_p$ results in decreasing length scales with minimal changes for the rotational structures. Results seem to asymptote near $Re_p>800$ and are consistent with $\langle L_L\rangle/D_H$.

 \hyperref[fig:L_ci]{Figure~\ref{fig:L_ci} (c)}, exhibits the size of vortical structures detected based on $\lambda_{ci}$ \cite{berdahl1993eduction,zhou1999mechanisms}, as well as $\lambda_{cr}/\lambda_{ci}$ criteria normalized by $D_H$. The length scales are determined based on the thresholds of $0.001\le\lambda_{ci}\le0.03$ and \textbf{$0<\lvert\lambda_{cr}/\lambda_{ci}\rvert\le2$}. The swirling structures show very large sizes ($L_{\lambda_{ci}}$) at $Re_p = 100$ but then decrease to the results of the other methods discussed earlier. The size of spiralling structures ($L_{\lambda_{cr}/\lambda_{ci}}$) tend to increase with $Re_p$ and appears to reach an asymptotic length of $0.2D_H$ near $Re_P \sim 1000$ again consistent with the other measures. This trend is consistent with what has been observed in Figures \hyperref[fig:lambda_cr]{\ref{fig:lambda_cr} (a)} to \hyperref[fig:lambda_cr]{\ref{fig:lambda_cr} (f)} where the vortex core sizes of non-planar, orbital structures increases with $Re_p$. This slight increase can be speculated as the spiral structures are elongated at the pore-scale. Also, the concentration of these spiral structures is larger as $Re_p$ increases, where the vortices are closer to each other passing through the image plane of visualization in a time-frozen flow field. The same phenomena in a FCC case based on the helicity of vortical structures is observed during the transition from inertial to turbulent flow regimes \cite{hill2002transition,he2019characteristics}. These results indicate that vortical sizes are constrained by, or scaled well, with the geometrical size of the pore. This feature is in agreement with Pore Scale Prevalence Hypothesis (PSPH) hypothesis \cite{nield1992convection, nield1991limitations,jin2015numerical}, where the turbulent structures are postulated to exist within pore-scales and not beyond.

 \begin{figure*}
 \centering
 \includegraphics[width=0.77\textwidth]{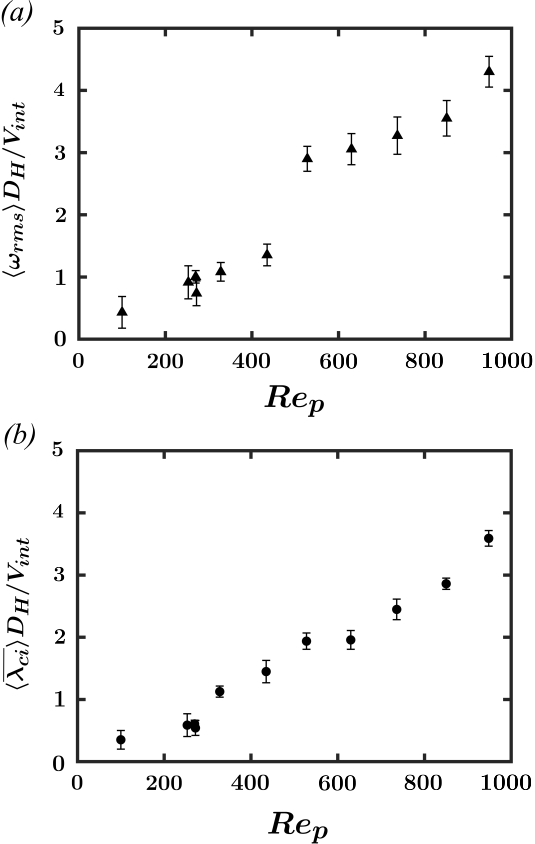}
 \caption{The variation of vortex strength versus $Re_p$ normalized by ($D_H/V_{int}$) (a) normalized vorticity rms $\langle \omega_{rms}\rangle D_H/V_{int}$, and (b) normalized time- and space-averaged vortex strength $\langle \overline{\lambda_{ci}}\rangle D_H/V_{int}$.}
\label{fig:omegarms}
 \end{figure*}

The spatio-temporal average of vortex strength is measured by both the rms vorticity and swirl strength, normalized by $D_H/V_{int}$ as shown in \hyperref[fig:omegarms]{Figure~\ref{fig:omegarms}} (where $V_{int}$ corresponds to each specific $Re_p$). The vorticity strength is seen to increase with $Re_p$ (\hyperref[fig:omegarms]{Figure~\ref{fig:omegarms} (a)}). Similarly, $\langle \overline{\lambda_{ci}}\rangle D_H/V_{int}$ shows an increasing trend with slightly lower values in \hyperref[fig:omegarms]{Figure~\ref{fig:omegarms} (b)}. The asymptotic level of the strength of vortices based on the inverse of the pore-averaged integral time scale is found for $Re_p$ larger than 1000 is shown by Patil and Liburdy (2013) \cite{patil2013turbulent}, to be $5.6V_{int}/D_H$. This value is fairly close with the value of strength found in the current study ($\sim 4-5V_{int}/D_H$).

The number of vortical structures identified increases with increasing the $Re_p$ as would be expected as shown in \hyperref[fig:count]{Figure~\ref{fig:count} (a)}. These results are the temporal average of the number of vortical structures detected based on $\lambda_{ci}$ and $\varGamma_{2}$ within an imaging plane with a given planar porosity $\Phi_{PL}$ (given in \hyperref[tab:cases]{Table~\ref{tab:cases}}). The results for the $\lambda_{ci}$ detection increase monotonically with $Re_p$ beyond $Re_p = 300$. Results based on different Large Eddy Scale (LES) filter sizes show different values, based on a Gaussian filter size of 3$\times$3 ($0.18D_H$) versus 5$\times$5 ($0.31D_H$). The smaller filter represents scales above 0.18$D_H$ corresponding to the asymptotic results of length scales illustrated in \hyperref[fig:L_ci]{Figure~\ref*{fig:L_ci}}, which is approximately 0.2$D_H$. This indicates that the number of both small and larger scales are increasing with $Re_p$. 

A spike is observed in \hyperref[fig:count]{Figure~\ref{fig:count} (a)} for $Re_p=267-270$ corresponding to PL-2, PL-3, and PL-4. These planes are located closely within the bed such that they share pores, and expectedly, vortical structures. PL-3 has higher $\Phi_{PL}$ (see \hyperref[tab:cases]{Table~\ref*{tab:cases}}) consistent with its larger average number of vortices compared with PL-2, and PL-4. The results based on $\varGamma_2$ detection (see \hyperref[fig:count]{Figure~\ref{fig:count} (b)}) are approximately the same order. Here, shear dominant ($\varGamma_{2} \le 2/\pi$) and rotational dominant ($\varGamma_{2} \ge 2/\pi$) values are delineated. Interestingly, the shear dominant number decrease, while rotationally dominant ocurrances increase. This is consistent with the idea of smaller structures developing within the pores further from the boundaries. The shear dominant vortices are the highest when $Re_p \le 300$.

\begin{figure*}
 \centering
 \includegraphics[width=0.75\textwidth]{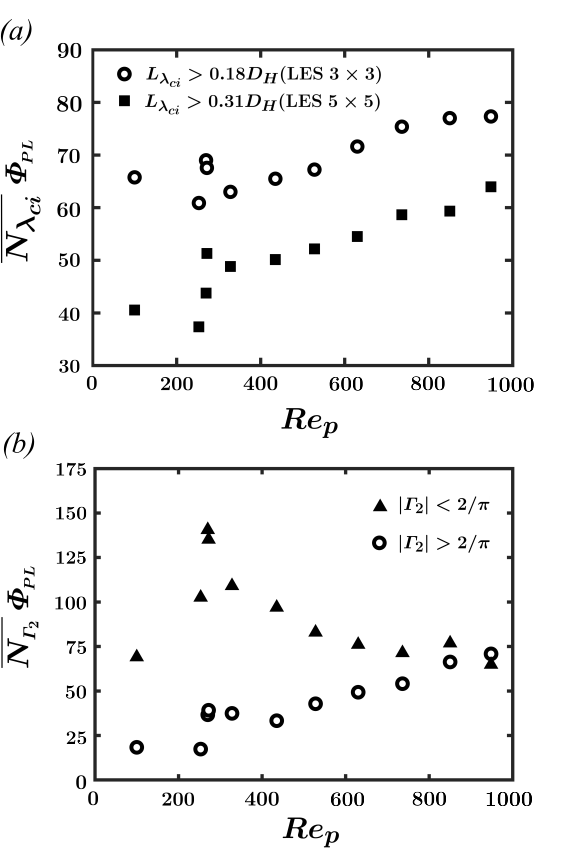}
 \caption{Temporal average of the number of detected vortical structures over all planes using (a) $\lambda_{ci}$ criterion with Gaussian filters for two kernel sizes of 3$\times$3 and 5$\times$5, and (b) $\Gamma_{2}$ for the shear-dominant votices ($\lvert \Gamma_{2} \rvert\leq 2/\pi$), and rotation-dominant vortices ($\lvert \Gamma_{2} \rvert\ge 2/\pi$).}
\label{fig:count}
 \end{figure*}

\subsection{Local scale of flow structures} \label{subsec:local} 
Pore-level scale estimates of vortical structures for the range of pore-scale Reynolds numbers ($Re_{\langle p\rangle}$) defined in \hyperref[sec:Introduction]{Section~\ref*{sec:Introduction}} are investigated in this section. These results emphasize local conditions and scaling. The local scales of the flow structures is based on each pore's value of length scale and velocity scale. Results are presented for integral length scales, vortex size using $\lambda_{ci}$, $\varGamma_{2}$, local time-scales using vortex strength, and number density. The hydraulic diameter of individual pores ($D_{H_{\langle p\rangle}}$) is found from the following:

\begin{equation}
D_{H_{\langle p\rangle}} = \frac{4A_{\langle p\rangle}}{\mathcal{P_{\langle \it{p}\rangle}}}
\label{eq:hydraulic} 
\end{equation}

\begin{figure*}
 \centering
 \includegraphics[width=1\textwidth]{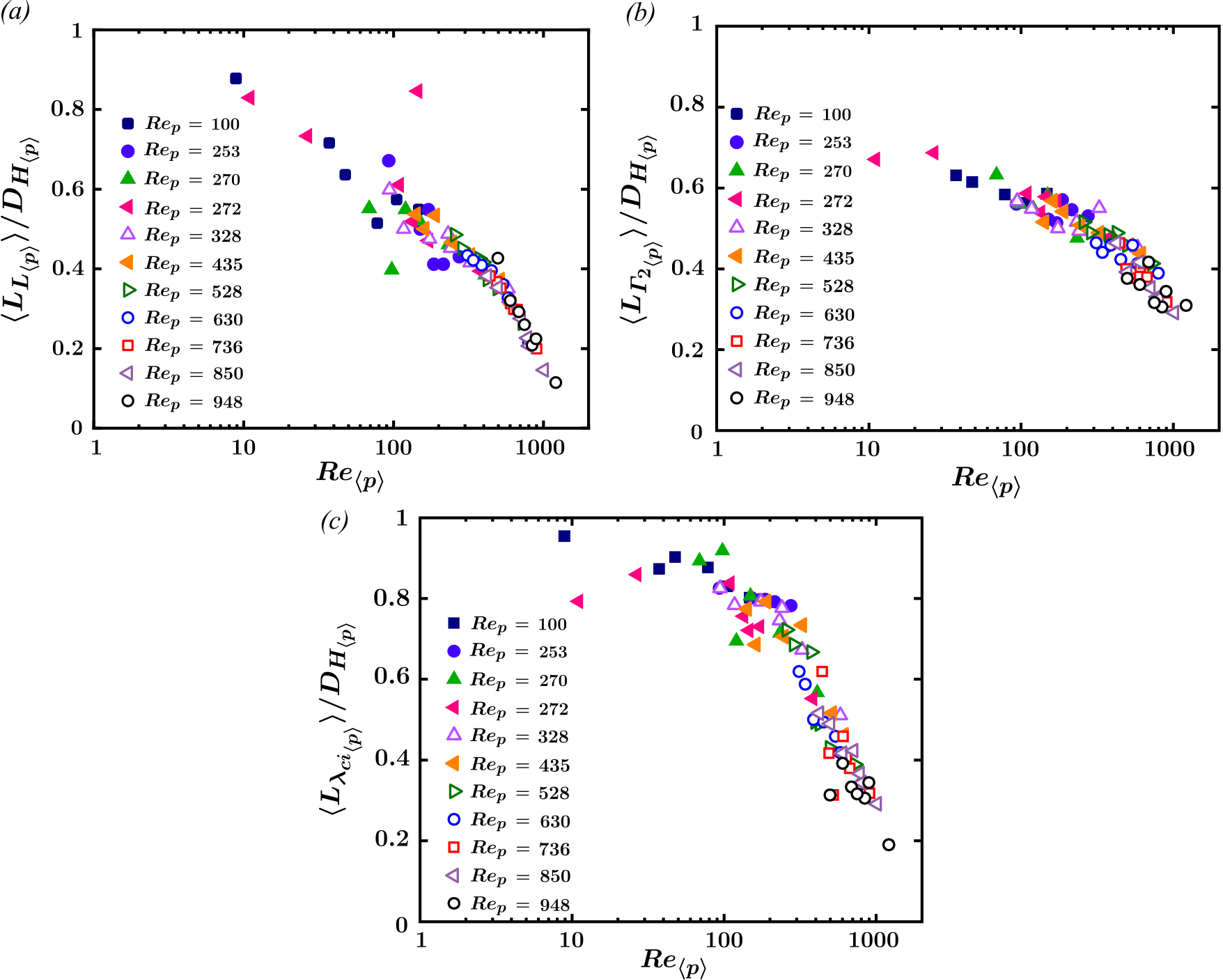}
 \caption{Variations of vortex size as a function microscopic Reynolds number ($Re_{\langle p\rangle}$) normalized by $D_{H_{\langle p\rangle}}$. (a) The variations of vortex size computed based on pore-averaged longitudinal integral length scale normalized by $ D_H$, (b) vortex size computed based on vortex boundary scheme ($\Gamma_{2}$), and (c) vortex size based on $\lambda_{ci}$ ($L_{\lambda_{ci}}$) criteria.}
\label{fig:local_int}
 \end{figure*}
 
\noindent where, $A_{\langle p\rangle}$ is the fluid area of an individual pore with boundaries indicated in \hyperref[fig:time-avg]{Figure~\ref{fig:time-avg}} (by excluding the solid areas of the beads), and $\mathcal{P_{\langle \it{p}\rangle}}$ is the wetted perimeter of the pore calculated by adding the solid boundary of pores (arc lengths of the pore walls) excluding the pore flow inlet and outlet areas.

   In \hyperref[fig:local_int]{Figure~\ref{fig:local_int}}, the overall size of vortical structures versus their corresponding $Re_{\langle p\rangle}$ is given. \hyperref[fig:local_int]{Figure~\ref{fig:local_int} (a)} is the pore-averaged longitudinal Eulerian integral length scale $\langle L_{L_{\langle p\rangle}}\rangle$ normalized by local pore-scale hydraulic diameter $D_{H_{\langle p\rangle}}$. Results show a decreasing trend with increasing $Re_{\langle p\rangle}$ on a log scale. For $Re_{\langle p\rangle}$ beyond approximately 100 data show a well-established and strong trend indicate that transition has occurred. \hyperref[fig:local_int]{Figure~\ref{fig:local_int} (b)} shows the length scales based on $\varGamma_{2}$ that also decreases very smoothly with $Re_{\langle p\rangle}$. Results found using $\lambda_{ci}$ is shown in \hyperref[fig:local_int]{Figure~\ref{fig:local_int} (c)}, with a smaller trend as found for the other two methods but indicates a significantly greater rate of decay. These later results indicate larger structures at the lower values of $Re_{\langle p\rangle}$ but similar size of flow structures at the larger $Re_{\langle p\rangle}$ values. Given the wide variation of pore geometries and sizes as well as resultant $Re_{\langle p\rangle}$ values, these data show a well-established trend based on individual pore condition, but not the bed averaged results.

 \begin{figure*}
 \centering
 \includegraphics[width=0.8\textwidth]{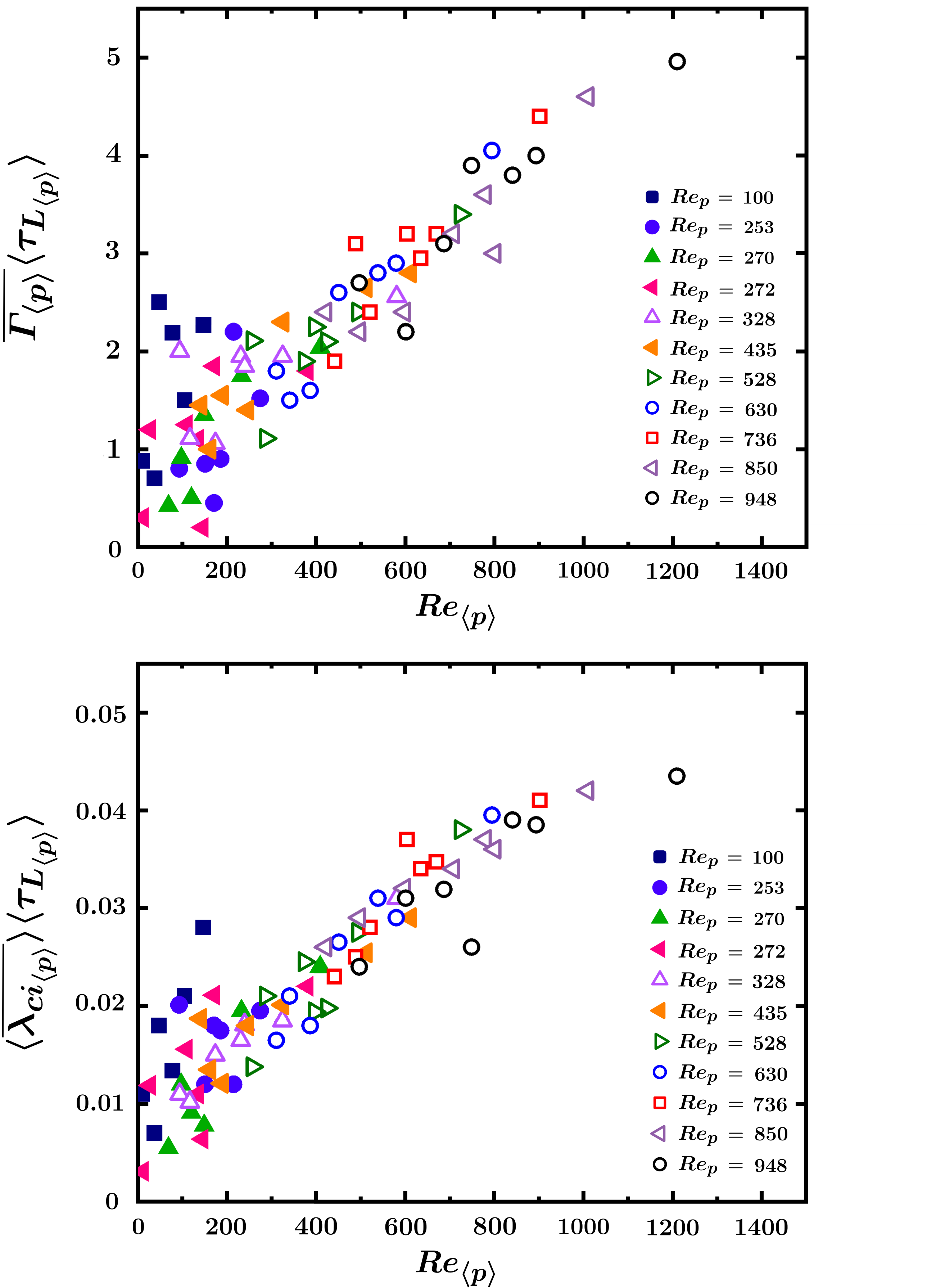}
 \caption{The variation of time averaged pore-scale vortex strength as a function pore-scale Reynolds number ($Re_{\langle p\rangle}$) non-dimensionalized by pore-averaged longitudinal time scale $\langle \tau_{L_{\langle p\rangle}}\rangle$. (a) Time averaged circulation $\overline{\varGamma_{\langle p\rangle} \langle} \tau_{L_{\langle p\rangle}}\rangle$, and (b) time and pore-averaged vortex strength $\langle \overline{\lambda_{ci_{\langle p\rangle}}}\rangle \langle \tau_{L_{\langle p\rangle}}\rangle$.}
\label{fig:Gamma}
 \end{figure*}

\begin{figure}[htbp]
             \centering
        \includegraphics[scale=0.32]{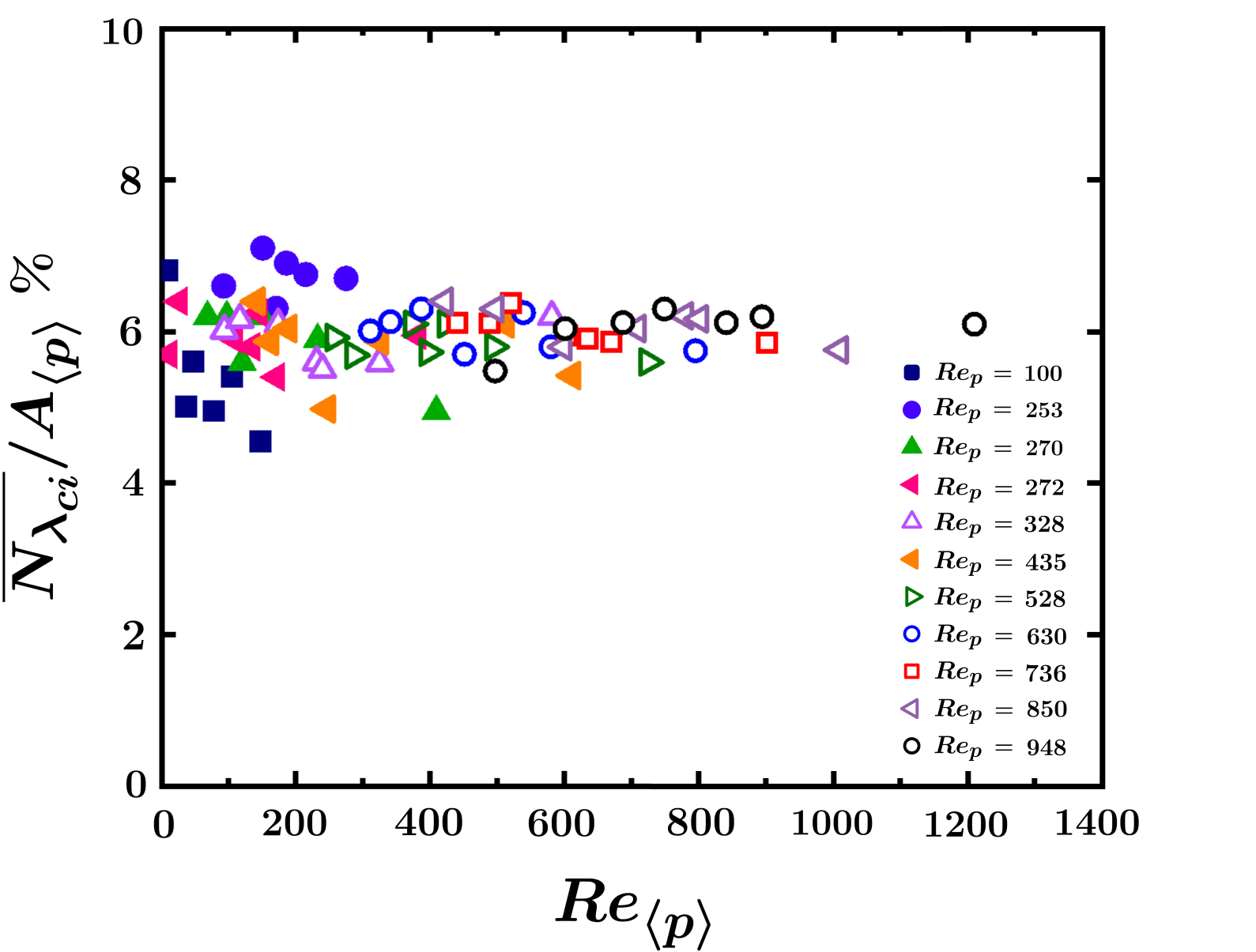}
        \caption{Temporal average of number of detected vortical structures in the entire bed based on $\lambda_{ci}$ criterion representing the number density of pore-scale vortical structures normalized by the area of pores ($A_{\langle p\rangle}$).}
        \label{fig:count-L}
\end{figure}

\hyperref[fig:Gamma]{Figure~\ref{fig:Gamma}(a)} shows the non-dimensional strength of vortical structures evaluated for the range of local pore Reynolds numbers ($Re_{\langle p\rangle}$) using individual pore scaling. The total non-dimensional time-averaged circulation within a pore, and the is evaluated. The variation of circulation in each pore is dependent upon the pore-scale vorticity as plotted in  \hyperref[fig:Gamma]{Figure~\ref{fig:Gamma}(a)}, where the circulation is calculated based on the sum of vorticity in each pore. The flow structures captured by circulation are originated from vorticity that included straining motions, pure shearing, and rotational structures \cite{kolavr2007vortex}. The dominance of shear structures for low $Re_{\langle p\rangle}$ ($Re_{\langle p\rangle}$<300) contributes to an increase in variation of circulation, while the rotation is the major contributor for a larger circulation in higher $Re_{\langle p\rangle}$ ($Re_{\langle p\rangle}$ > 300). As flow transitions to turbulence, the value of $\overline{\varGamma_{\langle p\rangle}}\langle\tau_L \rangle$ is expected to increase with $Re_{\langle p\rangle}$ due to the larger vorticity created by velocity gradients in tortuous random pores. There is significant scatter for smaller $Re_{\langle p\rangle}$ ($Re_{\langle p\rangle}$ < 300), which is thought to be a result of transition to turbulence at the pore inflow conditions, and/or three dimensional effects. However, the trend shows less scatter at larger $Re_{\langle p\rangle}$s indicating pores with higher local vorticity. The pore-scale circulation of vortical structures follows a linear trend in $Re_{\langle p\rangle}>300$. For this range ($Re_{\langle p\rangle}>300$), the global $Re_p$ is also large indicative of transition in the vicinity of $Re_{\langle p\rangle}\approx300$.

\hyperref[fig:Gamma]{Figure~\ref{fig:Gamma}(b)}, based on non-dimensional time- and pore-averaged vortex strength ($\overline{\lambda_{ci_{\langle p\rangle}}} \langle\tau_L \rangle$), illustrates the same trend as $\overline{\varGamma_{\langle p\rangle}}\langle\tau_L \rangle$. However, the rate of increasing $\overline{\lambda_{ci_{\langle p\rangle}}}\langle\tau_L \rangle$ are seen to be reduced after approximately $Re_{\langle p\rangle}>800$. This asymptotic trend is consistent with what has been observed by the number of vortical structures globally at $Re_p>800$ in \hyperref[fig:count]{Figure~\ref{fig:count} (a)}, where the trends for both scales ($3\times3$ and $5\times5$) flattened. Unlike $\overline{\varGamma_{\langle p\rangle}}\langle\tau_L \rangle$, the variations of $\overline{\lambda_{ci_{\langle p\rangle}}} \langle\tau_L \rangle$ is lower at $Re_{\langle p\rangle} < 300$. 

The pore-scale variations of $\overline{\varGamma_{\langle p\rangle}}\langle\tau_L \rangle$ and $\overline{\lambda_{ci_{\langle p\rangle}}} \langle\tau_L \rangle$ shown in \hyperref[fig:Gamma]{Figure~\ref{fig:Gamma}} represent the consistent behavior of the pore-scale vortical structures analogous to the global strength of vortices shown in \hyperref[fig:omegarms]{Figure~\ref{fig:omegarms}}. As a visual comparison, vorticity near the surface of spheres at lower $Re_{\langle p\rangle}$s are shear dominant with large strength due to the velocity gradient as shown by (i) and (ii) in \hyperref[fig:Lambda]{Figure~\ref{fig:Lambda-time}}. Therefore, the variations of $\overline{\lambda_{ci_{\langle p\rangle}}} \langle\tau_L \rangle$ is mostly influenced by those shear dominated vortical structures. Similarly, strong vorticity regions at the inflow of a pore shown in \hyperref[fig:vort-time]{Figure~\ref{fig:vort-time}} is evidence of large variations of $\overline{\varGamma_{\langle p\rangle}}\langle\tau_L \rangle$ at low $Re_{\langle p\rangle}$s.

The percent of local time-averaged number of vortical structures normalized by the local pore area ($A_{\langle p\rangle}$) versus $Re_{\langle p\rangle}$ is shown in \hyperref[fig:count-L]{Figure~\ref{fig:count-L}}. As shown, the ratio of number of vortical structures relative to the pore size is invariant with respect to $Re_{\langle p\rangle}$. The concentration of vortices are seen to be independent of the pore Reynolds number. Based on this, the overall number density of vortical structures is found to be a function of pore size. The number of detected vortices passing throughout each pore during the transition regime shows an invariance with respect to the size (area) of the pore. This seem to be the result of two possible features: pore-confinement, and the inertial effects. The level of confinement in random packed beds can control the possibility of pore-scale development of vortical structures. That is, pores with smaller sizes are reluctant to attract rotating structures from the mean flow and/or neighboring pores or they are small enough for the vortical structures to transport downstream. Moreover, pores with larger sizes are more prone to be influenced by the mean flow inertia that possibly increases the opportunity for the pore to develop shear effects. These conclusions are analogous to the geometry effects in high porosity random packed beds reported in simulations performed by Hill et al. (2001) \cite{hill2001moderate}. Their findings shows the wavelengths of vortical flow structures (flow instabilities) are damped. Hence, vortical structures are less apt to be observed in highly confined pores.
\section{Conclusion}\label{sec:conclusion}

High fidelity, time resolved particle image velocimetry were performed to identify vortical flow structures within a randomly packed porous media in six planes throughout the bed at global Reynolds numbers from 100 to 948. The effects of changing the Reynolds numbers as well as bed and pore geometry effects are investigated based on the vorticity distributions and vortical structures to evaluate size, strength (frequency), and number density using both macroscopic (global), and pore-scale (local) scales. Emphasis is placed on vortical structure scale evolution during transition and to scrutinize the use of the pore-scale Reynolds number to observe trends during transition in a random bed. The main observations are summarized below.

The mean flow influences the pore-scale flow due to tortuous characteristic of the random packing and generates wake-structures behind spheres, dead-zones, re-circulating vortical patterns, or elbow-like flows produced based on the geometry of pore, and flow acceleration, or deceleration of flow. These mean flow structures are observed not to exceed the limit of the local pore-level length scale orpore-scale hydraulic diameter ($D_{H_{\langle p\rangle}}$). 

Vortical structures tend to demonstrate a steady behavior for lower values of $Re_p$ ($<300$). These quasi-steady flow structures are shear dominant and generated near the surfaces of spheres and enlarge to decelerate the mean flow. For larger values of $Re_p$ ($>300$), structures near the surface of spheres are broken up into smaller vortices that are advected downstream within pores. 

Globally, the decay rate of the non-dimensional size of vortical structures versus the global $Re_p$ is similar to that observed when evaluated versus the pore-scale $Re_{\langle p\rangle}$. Some of these structures are rotational that tend not to vary in size with variation of $Re_p$, however, shear regions are shown to be reduced in size with increasing $Re_p$. The size of vortical structures using multiple techniques reach asymptotic values of $0.2-0.25D_H$ at  $Re_p \sim 1000$. The level of compactness of vortical structures is reduced with increasing $Re_p$, hence, the size and number of vortical structures with spiral behavior is larger for $Re_p > 500$. Locally, the size of vortical structures does not grow larger than the pore size or the global hydraulic diameter, hence it is in agreement with the PSPH hypothesis for the growth beyond the pore size. The dependence of integral length scale on the mean flow inertia is lower for small $Re_{\langle p\rangle}$. Additionally, larger variations occur of the size of vortices for $Re_{\langle p\rangle} < 100$. 

The vortex strength linearly increases with increasing $Re_p$. Locally, larger variations of strength are observed for low $Re_{\langle p\rangle}$. Moreover, the dominance of shear structures at lower $Re_{\langle p\rangle}$ contributes to an increase in the variation of  circulation, however, the main reason for the increase in variation of circulation at larger $Re_{\langle p\rangle}$ is due to rotational vortices. Additionally, the pore-scale rate of growth in vortex strength is consistent with two different techniques implemented (i.e. $\overline{\varGamma_{\langle p\rangle}}$ and $\langle\overline{ \lambda_{ci_{\langle p\rangle}}}\rangle$), but the variation of vortex strength for $Re_{\langle p\rangle} < 300$ is slightly larger. It is speculated based on these observations, that the local geometry and confinement as well as three-dimensional flow in the pores have significant effect on the pore-scale strength of vortical structures.

Globally, the time-averaged number of vortical structures grows with increasing $Re_p$. Moreover, the total average number of vortical structures passing through each plane increases with the planar porosity. The shear dominant structures are reduced for increasing $Re_p$, whereas the rotational structures grow in number. Locally, it is shown that the larger the pore size, the more vortical structures are advected. Hence, the number of local vortices is seen to be influenced by the pore size during the transition.

The dependence of flow structures has been found to be correlated with the effects from local inertial in the flow which is in accordance with PSPH. Also, well-defined trends of transitional vortical scaling of size, strength and number density on individual pore has been observed. The scales are influenced by the pore geometry (size and curvature of interstices) of pores, and pore-scale Reynolds number. These results provide an alternate means to investigate the transition process particularly for random bed geometries. Findings of this study is an effort towards a better understanding of the evolution of the flow structures during transition in random packed beds.\\



\bibliography{aipsamp}
%
\end{document}